\documentclass[aps,superscriptaddress,nofootinbib,eqsecnum,11pt]{revtex4}
\usepackage[english]{babel}
\usepackage[utf8]{inputenc}
\usepackage{graphicx}   % need for figures
\usepackage{slashed}
\usepackage{verbatim}   % useful for program listings
\usepackage{color}      % use if color is used in text
\usepackage{subfigure}  % use for side-by-side figures
\usepackage{multirow}
\usepackage{hyperref}   % use for hypertext links, including those to external documents and URLs
\usepackage{float}
\usepackage{epsfig,rotating}
\usepackage{amsmath,amssymb}
\usepackage{dsfont}
\restylefloat{table}
\numberwithin{equation}{section}

\newcommand{\vx}{\vec{x}}

\newcommand{\vk}{\vec{k}}

\newcommand{\be}{\begin{equation}}
\newcommand{\ee}{\end{equation}}
\newcommand{\bea}{\begin{eqnarray}}
\newcommand{\eea}{\end{eqnarray}}

\newcommand{\mgk}{\mathcal{G}_{k}}
\newcommand{\mg}{\mathcal{G}}
\begin{document}
%\tableofcontents
\title{ Information loss in effective field theory: entanglement and thermal entropies.}

\author{Daniel Boyanovsky}
\email{boyan@pitt.edu} \affiliation{Department of Physics and
Astronomy, University of Pittsburgh, Pittsburgh, PA 15260}
\date{\today}

\begin{abstract}
Integrating out high energy degrees of freedom to yield a low energy effective field theory leads to a loss of information with a concomitant increase in entropy.
We obtain the effective field theory of a light scalar field interacting with heavy fields after tracing out the heavy degrees of freedom from the time evolved   density matrix.  The initial density matrix describes the light field  in its ground state and the heavy fields   in equilibrium at a common temperature $T$. For $T=0$, we obtain the reduced density matrix in a perturbative expansion, it reveals   an emergent mixed state as a consequence of the entanglement between light and heavy fields. We obtain the   effective action that determines the time evolution of the \emph{reduced} density matrix for the light field   in a non-perturbative Dyson resummation of one-loop correlations of the heavy fields.  The Von-Neumann \emph{entanglement entropy}   associated with the reduced density matrix is obtained for  the non-resonant and resonant cases in the asymptotic long time limit. In the non-resonant case the reduced density matrix displays an \emph{incipient} thermalization albeit with a wave-vector,  time and coupling dependent \emph{effective temperature} as a consequence of memory of initial conditions.  The entanglement entropy is time independent and is the \emph{thermal entropy} for this effective, non-equilibrium temperature. In the resonant case the light field fully \emph{thermalizes} with the heavy fields, the reduced density matrix looses memory of the initial conditions and the entanglement entropy becomes the \emph{thermal entropy} of the light field. We discuss the relation between the entanglement entropy ultraviolet divergences and renormalization.
\end{abstract}

%\keywords{}

\maketitle

\section{Introduction, Motivation and goals}\label{sec:intro}

Effective field theory describes physical phenomena below some energy scale or on large spatio-temporal scales and is widely used in  different fields. Such an effective description has now become a paradigmatic pillar in critical phenomena\cite{ma}, condensed matter\cite{fradkin}, hydrodynamics\cite{nicolis,polo}, particle and nuclear physics\cite{eft1,eft2,eft3,eft4,dawson} and cosmology\cite{weinberg,cheung,senatore,nuestroreviu}. At a fundamental level, an effective field theory emerges upon coarse graining, namely tracing or integrating out, high energy or short distance degrees of freedom or fluctuations on small time scales.  One implementation of the concept of coarse graining is the renormalization group approach \emph{a l\'a} Wilson\cite{gold}.

Consider a \emph{full} theory describing interacting low and high energy degrees of freedom, in principle this complete theory has \emph{all the information} about \emph{all}  degrees of freedom, namely all the correlation functions can in principle be obtained. Tracing or integrating out the high energy degrees of freedom to obtain an effective field theory for the low energy degrees of freedom leads, therefore, to a \emph{loss} of information. If the quantum state that describes the \emph{full} theory is a pure state, namely a zero entropy state, tracing out the heavy degrees of freedom yields a \emph{reduced} density matrix for the low energy degrees of freedom which typically describes a \emph{mixed} state. Therefore   integrating out high energy or short distance degrees of freedom   to yield an effective field theory in principle leads to a mixed state with non-vanishing entropy. The   Von-Neumann entropy  is a measure of  the loss of information in the process of coarse graining.  The concept of a reduced density matrix originally introduced in pioneering work on quantum Brownian motion\cite{feyver,leggett} is now at the heart of the description of quantum open systems\cite{breuer,zoller}. This effective   coarse grained description while ubiquitous in condensed matter and quantum information\cite{qinfo}, has also received  attention   in quantum field theory out of equilibrium\cite{calhubuk,calhu,flemhu,qbmboy}, and more recently in cosmology\cite{boycosmo,holman,shandera,hollowood,lim} and   particle physics\cite{braaten,deboni,agon,bala,teresi} with intriguing connections to the information paradox in black hole physics\cite{harlow}. The effective field theory  approach out of equilibrium yields a stochastic description for the low energy degrees of freedom\cite{calhubuk,qbmboy,calhu}, and as a framework for quantum open systems opens the possibility of   extending the emerging field of quantum thermodynamics\cite{ford} to the realm of quantum field theory. This possibility has been bolstered recently  with a quantum field theory  extension of fluctuation theorems, with fundamental connections to entropy production and information\cite{jar}.

\vspace{1mm}

\textbf{Motivation and goals:}

Since at a fundamental level an effective field theory emerges after integrating or tracing out high energy (or short distance) degrees of freedom and describes only  the low energy degrees of freedom, the reduction from the full field theory (whatever its origin may be) to the effective one entails a loss of information and a concomitant increase of the entropy. Our study is motivated by this observation and seeks to understand the Von-Neumann entropy associated with the effective field theory in a model   that includes the essential  features and allows us to draw more general conclusions. This study is a natural continuation of a previous one\cite{qbmboy} that established the relation between non-equilibrium effective field theory, stochastic field theory and the quantum master equation approach. We consider a light scalar field coupled to other heavy scalar fields considering that the light field is in its ground state initially, whereas the heavy fields are in  thermal equilibrium at a common temperature $T$, the limit $T\rightarrow 0$ projects the ground state in the heavy sector. We evolve the initial density matrix in time and trace over the heavy degrees of freedom obtaining the reduced density matrix for the light field, and obtain the Von-Neumann entropy associated with the reduced density matrix in the asymptotic long time limit. This entropy is \emph{different} from the geometric (area) entropy\cite{sorkin,srednicki,callan,cardy} but   both originate  from tracing out (coarse graining) degrees of freedom. We analyze the emergence of the entanglement entropy upon coarse graining both from a perturbative and non-perturbative point of view. The latter allows us to study in detail the effects of thresholds and the case in which the light field thermalizes with the heavy fields.

\section{The model and perturbative evolution}\label{sec:modelpert}
We consider the model of a light real scalar field $\phi$ of mass $m_0$ coupled to two real heavy scalar fields $\psi$ and $\chi$ of masses $M_1,M_2 $ respectively. The Lagrangian density is
\be \mathcal{L} = \frac{1}{2}\,\partial_\mu \phi \partial^\mu \phi-\frac{1}{2}m^2_0 \phi^2 + \frac{1}{2}\,\partial_\mu \psi \partial^\mu \psi-\frac{1}{2}M^2_1 \,\psi^2+ \frac{1}{2}\,\partial_\mu \chi \partial^\mu \chi-\frac{1}{2}M^2_2 \chi^2 - g \phi\,\psi\,\chi \,.\label{lag}\ee The bare mass $m_0$ will be renormalized, (see below),  with the renormalized mass denoted $m_r$, without loss of generality we consider the hierarchy $M_1 \geq M_2 \gg m_r$.

In this section we study the perturbative evolution of an initial factorized \emph{state} at zero temperature to highlight several of the main conceptual aspects, relegating to the next section the study of the evolution of an initial \emph{density matrix} non-perturbatively, including thermal effects. We take  the initial state to be
\be |\Psi(0)\rangle = |0\rangle_\phi\otimes|0\rangle_\psi\otimes|0\rangle_\chi \,. \label{inivac}\ee
The time evolution of this state  up to second order  in a perturbative expansion in the interaction picture  is given   by
\be |\Psi(t)\rangle = |\Psi(0)\rangle + |\Psi^{(1)}(t)\rangle +|\Psi^{(2)}(t)\rangle  + \cdots \label{timeevol} \ee where
\bea |\Psi^{(1)}(t)\rangle  & =  &  -i\int_0^t H_I(t_1) dt_1 \,|\Psi(0)\rangle \label{state1}\\
|\Psi^{(2)}(t)\rangle  & =  &  (-i)^2\int_0^t \int_0^{t_1}\,H_I(t_1)H_I(t_2) \, dt_1 \, dt_2 \,|\Psi(0)\rangle  = -i\int_0^t H_I(t_1)\,|\Psi^{(1)}(t_1)\rangle ~  dt_1 \label{state2}\eea
and
\be H_I(t) = g^2 \int d^3x \,\phi(\vec{x},t) \,\psi(\vec{x},t)\,\chi(\vec{x},t) \label{HI}\ee is the interaction Hamiltonian in the interaction picture. With the time evolved state we obtain the \emph{reduced} density matrix by tracing the \emph{pure state} density matrix $|\Psi(t)\rangle \langle \Psi(t)|$ over the $\psi,\chi$ fields, namely,
\be \rho_r(t) = \mathrm{Tr}_{\psi,\chi} \big(|\Psi(t)\rangle\langle\Psi(t)|\big)\,. \label{redvac}\ee The first  and second order states $|\Psi^{(1,2)}\rangle$ are obtained from the free field expansions for the various fields, for example
 \be \phi(\vec{x},t)=  \frac{1}{\sqrt{V}}\,\sum_{\vec{k}} \frac{1}{\sqrt{2E_\phi(k)}}\, \Big[a_{\vec{k}} \, e^{-ikx}+ a^{\dagger}_{\vec{k}}\,e^{ikx}\Big] \,, \label{friex}\ee  and similarly for the other fields. We find
 \be |\Psi^{(1)}(t)\rangle = \sum_{\vec{k},\vec{q}} C^{(1)}(\vec{k},\vec{q};t)\,|1_{\vec{k}}\rangle_{\phi}\otimes |1_{\vec{q}}\rangle_{\psi} \otimes
 |1_{\vec{p}}\rangle_{\chi} ~~;~~ \vec{p} = -\vec{q}-\vec{k}\,, \label{Psi1t}\ee this is an \emph{entangled} multiparticle state of the light and heavy fields.

 In second order, there
 are several contributions obtained from the second equality in eqn. (\ref{state2}), however only two of these contribute to the reduced density matrix (\ref{redvac}) up to second order: a) annihilate all particles from $|\Psi^{(1)}\rangle$ returning to the full vacuum state $|\Psi{(0)}\rangle$, or b) create another $\phi$ particle annihilating the single particle states of $\psi,\chi$, yielding
 \be |\Psi^{(2)}(t)\rangle = C^{(2)}_a (t) \,|\Psi(0)\rangle + \sum_{\vec{k}} C^{(2)}_b(\vec{k},t) \,|1_{\vec{k}},1_{-\vec{k}}\rangle_{\phi}\otimes |0\rangle_{\psi} \otimes |0\rangle_{\chi} +\cdots \,,\label{psi2} \ee the dots stand for multiparticle states that will not contribute to the reduced density matrix up to second order in the coupling $g$. This second order state describes \emph{correlated pairs} of light particles.  The coefficients $C^{(n)}$ are of order $g^n$ they can be obtained straightforwardly but we will not use their explicit expressions,  our purpose in this section is to highlight the nature of the reduced density matrix, relegating to the next section a more detailed study of the reduced density matrix in the more general case.

\begin{figure}[ht!]
\begin{center}
\includegraphics[height=4in,width=4.5in,keepaspectratio=true]{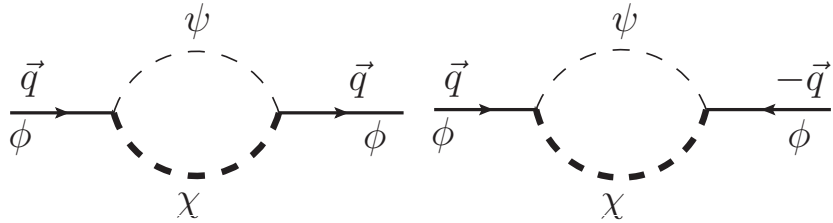}
\caption{ The second order contributions to the reduced density matrix.  }
\label{fig:rhosecord}
\end{center}
\end{figure}

 Up to second order the reduced density matrix reads
 \be \rho_r(t) = \rho_0(t) |0\rangle\langle 0| + \sum_{\vec{q}}\Bigg[\rho_1(\vec{q};t)|1_{\vec{q}}\rangle\langle 1_{\vec{q}}| +\rho_2(\vec{q};t) |1_{\vec{q}},1_{-\vec{q}}\rangle\langle 0| + \rho^*_2(\vec{q};t) |0\rangle\langle 1_{\vec{q}},1_{-\vec{q}}|  \Bigg] + \cdots \,.\label{rhosecor}\ee The explicit form of the  reduced density matrix elements $\rho_{0,1,2}$ are obtained from the coefficients $C^{(n)}$, they are not needed for the purpose of our arguments in this section. Although the reduced density matrix (\ref{rhosecor}) \emph{looks} like a mixed state, it is not a priori obvious that it is. Is it possible to find a state $|\alpha(t)\rangle$ so that

 \be |\alpha (t)\rangle \langle \alpha(t)| \stackrel{?}{=} \rho_r(t) \,,\label{pure}\ee if so the density matrix describes a pure, not a mixed state. Informed by the form of $\rho_r(t)$ in terms of states of single particles and correlated pairs  let us write generically up to second order
 \be |\alpha(t)\rangle = \alpha_0(t)|0 \rangle + \sum_{\vec{q}}\Big[ \alpha_1(\vec{q},t)|1_{\vec{q}}\rangle + \alpha_2(\vec{q}, t) |1_{\vec{q}},1_{-\vec{q}} \Big] \,,\label{alfastate}\ee where the coefficients $\alpha_{n}(t) \propto g^{n}$. Comparing  $|\alpha (t)\rangle \langle \alpha(t)|$ to  $\rho_r(t)$ eqn. (\ref{rhosecor}) we see that we can identify $|\alpha_0(t)|^2 = \rho_0(t)~~;~~ \alpha_2(\vec{q},t) = \rho_2(\vec{q},t)$, however the product state
$|\alpha (t)\rangle \langle \alpha(t)|$ yields  a term of the form $\sum_{\vec{q}}\big( \alpha_1(\vec{q},t) |1_{\vec{q}}\rangle \langle 0| + \mathrm{h.c.}\big)$, which is of order $g$, furthermore, there would also be second order contributions of the form $\sum_{\vec{q}\neq \vec{q}^{\,'}}   \big( \alpha_1(\vec{q},t)\,\alpha^*_1(\vec{q}^{\,'},t) |1_{\vec{q}}\rangle \langle 1_{\vec{q}^{\,'}}| + \mathrm{h.c.}\big)$. \emph{Neither} of these contributions is present in $\rho_r(t)$, in particular there is no term of $\mathcal{O}(g)$ in $\rho_r(t)$ precisely because the trace over the heavy degrees of freedom requires \emph{pairs} of $\psi,\chi$ fields. The trace over the intermediate $\psi,\chi$ states  rules out a reduced state of the form $|1_{\vec{q}}\rangle \langle 0|$, and  forces $\vec{q} = \vec{q}^{\,'}$ by momentum conservation in single particle   states $|1_{\vec{q}}\rangle \langle 1_{\vec{q}^{\,'}}|$. This analysis leads us to conclude that, indeed, $\rho_r(t)$ describes a mixed state. An interpretation of the second order contributions to $\rho_r(t)$ is depicted in fig. (\ref{fig:rhosecord}).

 The main results  of this perturbative study are:

\textbf{a:)} The reduced density matrix is the \emph{effective} dynamical description of the light degrees of freedom and describes a \emph{mixed state} containing both single particles and correlated pairs of particles. The matrix elements are a direct consequence  of the \emph{entanglement} between the light   and   heavy fields the latter being traced over.

\vspace{1mm}

\textbf{b:)} Up to second order, the mixed nature of the reduced density matrix is revealed in the contribution from the single particle states $|1_{\vec{q}}\rangle \langle 1_{\vec{q}}|$. This aspect will be important in the discussion of the origin of entropy in the time evolution of the density matrix discussed in the next section.

\vspace{1mm}

    \textbf{c:)} Up to second order and restricting to the one and two (pair) particle states, the eigenvalues of the reduced density matrix are: $p_1 = 1+ \mathcal{O}(g^2); p_2 \simeq \mathcal{O}(g^2)$. Therefore we expect   the  Von-Neumann \emph{entropy} $S = -\sum_n p_n \ln(p_n)$ to be  of order $S    \simeq g^2 \,\ln(1/g^2)$. This entropy is a consequence of the entanglement between the light and heavy fields and must be interpreted as an \emph{entanglement entropy}\cite{cardy}. In comparing the pure state $|\alpha (t)\rangle \langle \alpha(t)|$ to  $\rho_r(t)$ eqn. (\ref{rhosecor}) it is clear that the entanglement entropy describes the loss of information in the states that are missing from $\rho_r(t)$. The \emph{kinematic} entanglement between the light and heavy fields prevents these (missing) states from appearing in the reduced density matrix after tracing   the heavy fields.

\section{Time evolution of the density matrix: effective action}

In the previous section we provided a perturbative interpretation of the reduced density matrix and the origin of the entanglement entropy from the coarse graining process, we now obtain the time evolution of the full density matrix from the effective action up to second order in the coupling.

It is argued below that the reduced density matrix obtained from the effective action corresponds to a \emph{non-perturbative} Dyson resummation of one-loop diagrams.

In order to simplify notation in the analysis below, we collectively define the heavy fields in terms of a doublet
\be h \equiv (\psi, \chi)\,, \label{heavy}\ee writing the total Lagrangian density and Hamiltonian as
\bea \mathcal{L}(\phi,h) & = &  \mathcal{L}_{0,\phi}(\phi)+\mathcal{L}_{0,h}(h) + \mathcal{L}_I(\phi,h) \label{lagfih} \\
H & = & H_{0,\phi}+H_{0,h}+H_I ~~;~~ H_I = g\int d^3x \phi(\vec{x})\psi(\vec{x})\chi(\vec{x}) \,, \label{Hfih}\eea where $\mathcal{L}_{0}, H_{0}$ refer to the free field Lagrangian density and Hamiltonian respectively.

Consider the initial density matrix at a time
$t=0$ to be of the form
\begin{equation}
 {\rho}(0) =  {\rho}_{\phi}(0) \otimes
 {\rho}_{h}(0) \,.\label{inidensmtx}
\end{equation}

We will consider that initially the $\phi$ field is in its vacuum state,
\be \hat{\rho}_{\phi}(0) = (|0\rangle \langle 0|)_\phi \,, \label{rhofini}\ee and the initial density matrix of the $h=(\psi,\chi)$ fields will be taken to
describe a statistical ensemble in thermal equilibrium at a common temperature
$T=1/\beta$, namely

\begin{equation}\label{rhochi}
\hat{\rho}_{h}(0) = \frac{e^{-\beta H_{0,  h}}}{\mathrm{Tr}_h \, e^{-\beta H_{0, h}}}\,.
\end{equation}   The zero temperature limit projects the vacuum state for the heavy fields, in this limit we will establish a correspondence with the perturbative analysis of the previous section.

In the Schroedinger representation (field basis) the matrix elements of $\rho_\phi(0),\rho_h(0)$ are given by
\begin{equation}
\langle \phi | {\rho}_{\phi}(0) | \phi'\rangle =
\rho_{\phi,0}(\phi ,\phi')~~;~~\langle h | {\rho}_{h}(0) | h'\rangle =
\rho_{h,0}(h ;h')\,,
\end{equation}   this is a \emph{functional} density matrix as the fields have spatial arguments.
This initial density matrix  will evolve  out of
equilibrium   since it does not commute with the full interacting Hamiltonian,
\be \rho(t) = U(t)\,\rho(0) \, U^{-1}(t) ~~;~~ U(t)= e^{-iHt}\,, \label{rhototoft}\ee where the total Hamiltonian is given by eqn. (\ref{Hfih}). The matrix elements of the time evolved density matrix in the field basis are given by
\bea   \rho(\phi_f,h_f;\phi'_f,h'_f;t) & = &      \langle \phi_f;h_f|U(t)\hat{\rho}(0)U^{-1}(t)|\phi'_f;h'_f\rangle \nonumber \\
& = & \int D\phi_i Dh_i D\phi'_i Dh'_i ~ \langle \phi_f;h_f|U(t)|\phi_i;h_i\rangle\,\rho_{\phi,0}(\phi_i;\phi'_i)\,\rho_{h,0}(h_i;h'_i)\,\nonumber \\
 & \times &  \langle \phi'_i;h'_i|U^{-1}(t)|\phi'_f;h'_f\rangle \,. \label{evolrhot}\eea The $\int D\phi$ etc, are functional integrals where the spatial arguments have been suppressed. The matrix elements of the time evolution forward and backward can be written as path integrals, namely
 \bea   \langle \phi_f;h_f|U(t)|\phi_i;h_i\rangle  & = &    \int \mathcal{D}\phi^+ \mathcal{D}h^+\, e^{i \int d^4 x \mathcal{L}[\phi^+,h^+]}\label{piforward}\\
 \langle \phi'_i;h'_i|U^{-1}(t)|\phi'_f;h'_f\rangle &  =  &   \int \mathcal{D}\phi^- \mathcal{D}h^-\, e^{-i \int d^3 x \mathcal{L}[\phi^-,h^-]}\label{piback}
 \eea with the shorthand notation
 \be \int d^4 x \equiv \int_0^t dt \int d^3 x \,,\label{d4xdef}\ee
 and $ \mathcal{L}[\phi,h] $ is given by (\ref{lagfih}). The   boundary conditions on the path integrals are
  \bea     \phi^+(\vec{x},t=0)=\phi_i(\vec{x})~;~
 \phi^+(\vec{x},t)  &  =  &   \phi_f(\vec{x})\,,\nonumber \\   h^+(\vec{x},t=0)=h_i(\vec{x})~;~
 h^+(\vec{x},t) & = & h_f(\vec{x}) \,,\label{piforwardbc}\\
     \phi^-(\vec{x},t=0)=\phi'_i(\vec{x})~;~
 \phi^-(\vec{x},t) &  = &    \phi'_f(\vec{x})\,,\nonumber \\   h^-(\vec{x},t=0)=h'_i(\vec{x})~;~
 h^-(\vec{x},t) & = & h'_f(\vec{x}) \,.\label{pibackbc}  
 \eea

The field variables $\phi^\pm, h^\pm$ along the forward ($+$) and backward ($-$) evolution branches are recognized as those necessary for the Schwinger-Keldysh\cite{schwinger,keldysh,maha,calhubuk} closed time path approach to the time evolution of a density matrix.

The reduced density matrix for the light field $\phi$ is obtained by tracing over the fields  $h = \psi;\chi$, namely

\be \rho^{r}(\phi_f,\phi'_f;t) = \int Dh_f \,\rho(\phi_f,h_f;\phi'_f,h'_f=h_f;t) \,,\label{rhored} \ee we find
\be \rho^{r}(\phi_f,\phi'_f;t) = \int D\phi_i   D\phi'_i  \,  \mathcal{K}[\phi_f,\phi'_f;\phi_i,\phi'_i;t] \,\rho_\phi(\phi_i,\phi'_i;0)\,. \ee
The propagating  kernel $\mathcal{K}$ is given by
\be \mathcal{K}[\phi_f,\phi_i;\phi'_f,\phi'_i;t] =  \int  \mathcal{D}\phi^+ \, \int \mathcal{D}\phi^- \, e^{i  \int d^4x \left[\mathcal{L}_0[\phi^+]-\mathcal{L}_0[\phi^-]\right]}\,e^{i\mathcal{F}[\phi^+;\phi^-]} \ee with
the following boundary conditions on the forward ($\phi^+$) and backward  ($\phi^-$) path integrals
\bea &  &   \phi^+(\vec{x},t=0)=\phi_i(\vec{x})~;~
 \phi^+(\vec{x},t)  =   \phi_f(\vec{x}) \nonumber \\
&  &   \phi^-(\vec{x},t=0)=\phi'_i(\vec{x})~;~
 \phi^-(\vec{x},t)  =   \phi'_f(\vec{x}) \,.\label{bcfipm}\eea   $\mathcal{F}[\phi^+;\phi^-]$ is the \emph{influence action}, arising from the trace over the heavy fields,   given by
 \be  e^{i\mathcal{F}[\phi^+;\phi^-]}   =   \int Dh_i  Dh'_i Dh_f  \int \mathcal{D}h^+  \mathcal{D}h^- \, e^{i  \int d^4x \left[\mathcal{L}_0(h^+)+\mathcal{L_I}(\phi^+,h^+)\right]}~   e^{-i  \int d^4x \left[\mathcal{L}_0(h^-)+\mathcal{L_I}(\phi^-,h^-)\right]}\,\rho_{h}(h_i,h'_i;0) \,, \label{inffunc}\ee the boundary conditions on the path integrals are
 \be h^+(\vec{x},t=0)=h_i(\vec{x})~;~
 h^+(\vec{x},t)=h_f(\vec{x})~~;~~ h^-(\vec{x},t=0)=h'_i(\vec{x})~;~
 h^-(\vec{x},t)=h'_f(\vec{x})=h_f(\vec{x}) \,. \label{bcchis} \ee where the last equality reflects the trace over the $h \equiv \psi, \chi$ fields.

\begin{figure}[ht!]
\begin{center}
\includegraphics[height=1.5in,width=2in,keepaspectratio=true]{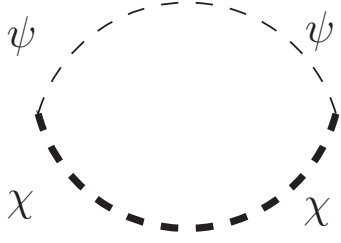}
\caption{ The generic correlation function $G(x_1-x_2)$ is a $\psi-\chi$ loop.  }
\label{fig:correlators}
\end{center}
\end{figure}

In the  path integral (\ref{inffunc}), $ \phi^\pm $ act  as an \emph{external c-number source} coupled to the composite operator $\psi(x) \chi(x) $ along each branch, therefore, it follows that
\be e^{i\mathcal{F}[\phi^+;\phi^-]} = \mathrm{Tr} \Big[ \mathcal{U}(t;\phi^+)\,\rho_h(0)\,  \mathcal{U}^{-1}(t;\phi^-) \Big]\,, \label{trasa}\ee where  $\mathcal{U}(t;\phi^\pm)$ is the   time evolution operator in the $\psi,\chi$ sectors in presence of \emph{external sources} $\phi^\pm$,  namely \be \mathcal{U}(t;J^+) = T\Big( e^{-i \int_0^t H_h[\phi^+(t')]dt'}\Big) ~~;~~
\mathcal{U}^{-1}(t;\phi^-) = \widetilde{T}\Big( e^{i \int_0^t H_h[\phi^-(t')]dt'}\Big) \label{timevchi}\ee  where
\be H_h[\phi^\pm(t)] = H_{0,h}+g\int d^3x \,\phi^\pm \,\psi\, \chi \,,\label{timevchiH}\ee and $\widetilde{T}$ is the \emph{anti-time evolution operator} as befits   evolution backward in time. The calculation of the influence action is facilitated by passing to the interaction picture for the Hamiltonian $H_h[\phi(t)]$, defining
\be  \mathcal{U}(t;\phi^\pm) = e^{-i H_{0,h}\,t} ~ \mathcal{U}_{ip}(t;\phi^\pm) \label{ipicture} \ee and the $e^{\pm i H_{0,h}\,t}$ cancel out in the trace in (\ref{trasa}). Now the trace can be obtained systematically in perturbation theory in $g$. Up to $\mathcal{O}(g^2)$ and with notation (\ref{d4xdef})  we find\cite{qbmboy}
\bea \mathcal{F}[\phi^+,\phi^-] & = &     \frac{i g^2 }{2} \int d^4x_1 \int d^4x_2 \Bigg\{ \phi^+(x_1)\,\phi^+(x_2)\,G^{++}(x_1-x_2)+ \phi^-(x_1)\,\phi^-(x_2)\,G^{--}(x_1-x_2) \nonumber \\
 & - & \phi^+(x_1)\,\phi^-(x_2)\,G^{+-}(x_1-x_2)- \phi^-(x_1)\,\phi^+(x_2)\,G^{-+}(x_1-x_2)\Bigg\}\,. \label{finF}\eea The   correlation functions are given by
 \begin{eqnarray}
&& G^{-+}(x_1-x_2) =   \langle
\psi(x_1) \psi(x_2)\rangle \langle
\chi(x_1) \chi(x_2)\rangle =   {G}^>(x_1-x_2) \,,\label{ggreat} \\&&  G^{+-}(x_1-x_2) =    \langle
\psi(x_2) \psi(x_1)\rangle \langle
\chi(x_2) \chi(x_1)\rangle  =   {G}^<(x_1-x_2)\,,\label{lesser} \\&& G^{++}(x_1-x_2)
  =
{ G}^>(x_1-x_2)\Theta(t_1-t_2)+ {  G}^<(x_1-x_2)\Theta(t_2-t_1) \,,\label{timeordered} \\&& G^{--}(x_1-x_2)
  =
{ G}^>(x_1-x_2)\Theta(t_2-t_1)+ {  G}^<(x_1-x_2)\Theta(t_1-t_2)\,,\label{antitimeordered}
\end{eqnarray} in terms of interaction picture fields, where
\be \langle (\cdots) \rangle = \mathrm{Tr}(\cdots)\rho_h(0)\,. \label{expec}\ee
Furthermore, for real scalar fields as considered here  it follows that
\be G^>(x_1-x_2) = G^<(x_2-x_1)\,. \label{iden}\ee These correlation functions describe one-loop contributions as shown in fig.(\ref{fig:correlators}) and are precisely the correlations that enter in the perturbative study in the previous section (see fig. \ref{fig:rhosecord}).

The \emph{effective action} out of equilibrium is given by
\be  {S}_{eff}[\phi^+,\phi^-] = \int^t_0 dt \int d^3 x \Big\{ \mathcal{L}_{0,\phi}(\phi^+)-\mathcal{L}_{0,\phi}(\phi^-)\Big\} +\mathcal{F}[ \phi^+ ;\phi^-] \,.\label{Leff} \ee

The influence action can be simplified and written solely in terms of the two independent correlation functions $G^>,G^<$ by the following steps\cite{qbmboy}:
\begin{itemize}
\item{For the contribution with $\phi^+(x_1)\phi^+(x_2)$: in the term $G^<(x_1-x_2)\Theta(t_2-t_1)$ (see eqn. (\ref{timeordered})) relabel $t_1 \leftrightarrow t_2$ and use the property (\ref{iden}).  }
\item{ For the contribution $\phi^-(x_1)\phi^-(x_2)$: in the term $G^>(x_1-x_2)\Theta(t_2-t_1)$ (see eqn. (\ref{antitimeordered})) relabel $t_1 \leftrightarrow t_2$ and use the property (\ref{iden}). }
    \item{ For the contribution  $\phi^+(x_1)\phi^-(x_2)$: multiply $G^<(x_1-x_2)$ by $\Theta(t_1-t_2)+\Theta(t_2-t_1)=1$ and in the term with $\Theta(t_2-t_1)$ relabel $ t_1 \leftrightarrow t_2$ and use the property (\ref{iden}). }
 \item{ For the contribution $\phi^-(x_1)\phi^+(x_2)$: multiply $G^>(x_1-x_2)$ by $\Theta(t_1-t_2)+\Theta(t_2-t_1)=1$ and in the term with $\Theta(t_2-t_1)$ relabel $t_1 \leftrightarrow t_2$ and use the property (\ref{iden}). }
\end{itemize}
Finally we find
\bea \mathcal{F}[\phi^+, \phi^-]  &  = &  i\,g^2\int d^3x_1 d^3x_2 \int^t_0 dt_1\,\int^{t}_0 dt_2\,\Bigg\{ \phi^+(\vx_1,t_1)\phi^+(\vx_2,t_2)\,G^>(x_1-x_2) \nonumber \\ & + &    \phi^-(\vx_1,t_1)\phi^-(\vx_2,t_2)\,G^<(x_1-x_2)
   -   \phi^+(\vx_1,t_1)\phi^-(\vx_2,t_2)\,G^<(x_1-x_2) \nonumber\\ & - &    \phi^-(\vx_1,t_1)\phi^+(\vx_2,t_2)\,G^>(x_1-x_2)\Bigg\}~\Theta(t-t_1) \label{Funravel}\eea where $G^{\lessgtr}$ are given by eqns. (\ref{ggreat},\ref{lesser}).  This is the general form of the influence function up to second order in the  coupling. The reduced density matrix is finally given by
   \be \rho^{r}(\phi_f,\phi'_f;t) = \int D\phi_i   D\phi'_i  \, \int  \mathcal{D}\phi^+   \mathcal{D}\phi^- \, e^{iS_{eff}[\phi^+,\phi^-;t]}  \,\rho_\phi(\phi_i,\phi'_i;0)\,, \label{rhorfin} \ee where the path integrals over $\phi^\pm$ are performed with  the boundary conditions (\ref{bcfipm}) and $S_{eff}$ is given by (\ref{Leff}).    Equation (\ref{rhorfin}) explicitly shows that the reduced density matrix evolves in time via the effective action.

The matrix elements of the initial density matrix $\rho_\phi(0)$ in the field basis are more conveniently written in terms of the spatial Fourier transform of the field in a spatial volume $V$,
\be \phi(\vec{x}) = \frac{1}{\sqrt{V}}\sum_{\vec{k}} \phi_{\vec{k}}~ e^{-i\vec{k}\cdot\vec{x}}\,. \label{FT} \ee The density matrix describing the vacuum state of free fields is given by
\be \rho_\phi(\phi_i,\phi'_i;0) = \prod_{\vk} ~ {N}_k ~e^{- \frac{\Omega_{0k}}{2} \, \big[\phi_{\vk,i}\,\phi_{-\vk,i}+ {\phi'}_{\vk,i}\,{\phi'}_{-\vk,i}\big]} ~~;~~ \Omega_{0k} = \sqrt{k^2+m^2_0}\,, \label{rhoinift}\ee the   frequency $\Omega_{0k}$ corresponds to the bare free field mass $m_0$, and the normalization factor $ {N}_k$ is fixed by the requirement that
\be \int D\phi_i ~ \rho_\phi(\phi_i,\phi'_i=\phi_i;0) =1 \,.\label{norma}\ee

We emphasize that while the reduced density matrix is obtained by tracing over the heavy degrees of freedom, the total density matrix evolves in time unitarily, this entails that $\mathrm{Tr} \rho(t) = \mathrm{Tr} \rho(0)$, this fact along with the normalization (\ref{norma}) yields
\be \int D\phi_f \, \rho^{r}(\phi_f,\phi'_f=\phi_f;t)  = 1 \,. \label{normafina}\ee This result is a consequence of unitary time evolution and normalization of the initial density matrix,  and will   be important in the discussion below.

\subsection{Correlation functions of heavy fields}
The correlation functions of the heavy fields $G^>,G^<$ can be written in terms of a spectral representation, for details see ref.\cite{qbmboy} and appendix (\ref{app:specdens}),

\be G^\lessgtr(x-x') = \frac{1}{V} \sum_{\vk} G^\lessgtr_k(t-t')\,e^{i\vk\cdot(\vx-\vx')}~~;~~ G^\lessgtr_k(t-t') = \int \frac{dk_0}{(2\pi)}\,\rho^\lessgtr(k_0,k)\,e^{-ik_0(t-t')}   \label{specrep}\ee with\cite{qbmboy}
\be \rho^>(k_0,k) = \rho(k_0,k)\big[1+n(k_0)] ~~;~~ \rho^<(k_0,k) = \rho(k_0,k) n(k_0) ~~;~~ n(k_0) = \frac{1}{e^{\beta k_0}-1} ~;~ \beta = 1/T \label{rhogl}\ee where $\rho(k_0,k)$ is the spectral density and $T$ is the common temperature of the heavy fields $\psi, \chi$. The spectral density for the case considered here of two scalar fields of masses $M_1,M_2$ at a common temperature $T$ has been obtained in ref.\cite{qbmboy}, it is given by
\bea \rho(k_0,k;T) & = &  \rho_{ld}(k_0,k;T)\,\Theta(-Q^2)+ \rho_d(k_0,k;T)\,\Theta((M_1-M_2)^2-Q^2)\,\Theta(Q^2)\nonumber \\ & + & \rho_{2p}(k_0,k;T)\,\Theta(Q^2-(M_1+M_2)^2) ~~;~~Q^2=k^2_0 - k^2 \,, \label{rho2chis}\eea where the explicit expressions for the different contributions are summarized in appendix (\ref{app:specdens}). Two important properties of the spectral density   will be relevant in the analysis:
\be \rho(k_0,k;T) = - \rho(-k_0,k;T) ~~;~~ \rho(k_0>0,k;T) >0 \,,\label{props}\ee

The contribution $\rho_{ld}(k_0,k;T)$ with support below the light cone ($Q^2 <0$) corresponds to the process of Landau damping. This is a medium dependent contribution that vanishes in the $T\rightarrow 0$ limit. It describes collisionless damping in a medium as a consequence of dephasing\cite{boylan}.

The contribution $\rho_d(k_0,k;T)$ also describes a process solely available in the medium. On the renormalized mass shell $Q^2 = m^2_r$, and  for $M_1 > M_2+ m_r$  it describes the \emph{decay}   $\psi  \rightarrow \chi \phi$ (since $M_1 > M_2 + m_r$).  This part of the spectral density also vanishes for $T\rightarrow 0$, however at $T\neq 0$ it \emph{has support on the renormalized mass shell of the light field $\phi$}. This term has a quantum kinetic interpretation\cite{qbmboy} in terms of the in-medium processes $\psi \leftrightarrow \chi \phi$. As discussed below, this contribution is responsible for the   \emph{thermalization} of the light field $\phi$ with the bath of heavy fields when the spectral density has support on the renormalized mass shell of the light field. We refer to the case with $M_1 > M_2 +m_r$, when the spectral density of the heavy fields has support on the (renormalized) mass shell of the light field as the \emph{resonant} case.

A relevant example of this scenario in particle physics is given by a charged current vertex in the standard model,  with the heavier field ($\Psi$) being the $W$ vector boson, the field $\chi$ being a charged lepton (for example the electron)  and the lightest field $\phi$ being a neutrino. At high temperature $W \leftrightarrow e\nu_e$  and the inverse process contributes  to  neutrino thermalization   as a consequence of detailed balance\cite{lellonu}.

The contribution $\rho_{2p}(k_0,k;T)$ corresponds to the usual two particle cut for $Q^2 > (M_1+M_2)^2$, it is the only contribution to the spectral density that does not vanish in the $T\rightarrow 0$ limit ($n(k_0) \rightarrow -\Theta(-k_0)$), where  it is given by (see appendix \ref{app:specdens})
\be \rho(k_0,k;T=0) = \frac{\mathrm{sign}(k_0) }{8\pi\,Q^2}\Bigg\{\Big[Q^2-(M_1-M_2)^2 \Big]\,\Big[Q^2-(M_1+M_2)^2 \Big] \Bigg\}^{\frac{1}{2}}\,\Theta\Big[Q^2-(M_1+M_2)^2\Big]\, ,   \label{zeroTrho2chi}\ee  The $\Theta$ function in (\ref{zeroTrho2chi}) corresponds to the two particle threshold.

\subsection{Path integral derivation of the effective action:}

Introducing the spatial Fourier transforms of $\phi^\pm$ as in (\ref{FT}), we introduce the variables
\be \Psi_{\vk}(t) = \frac{1}{2}\big(\phi^+_{\vk}(t) + \phi^-_{\vk}(t)\big)~~;~~ R_{\vk}(t)= \big( \phi^+_{\vk}(t)-\phi^-_{\vk}(t)\big) \label{WT} \ee with the boundary conditions
\bea \Psi_{\vk,i}   & \equiv & \Psi_{\vk}(0)  = \frac{1}{2}\big(\phi_{\vk,i} + \phi^{\,'}_{\vk,i}\big)~~;~~ \Psi_{\vk,f}   \equiv \Psi_{\vk}(t)  = \frac{1}{2}\big(\phi_{\vk,f} + \phi^{\,'}_{\vk,f}\big) \label{psibc}\\
 R_{\vk,i}  & \equiv &   R_{\vk}(0)=  \big(\phi_{\vk,i} - \phi^{\,'}_{\vk,i}\big)~~~;~~~ R_{\vk,f}\equiv R_{\vk}(t) = \big(\phi_{\vk,f} - \phi^{\,'}_{\vk,f}\big) \,.\label{Rbc} \eea The effective action becomes
 \bea  i S_{eff}  & = &   \sum_{\vk} \Bigg\{\int^t_0 dt_1 \,i \Big[ \dot{R}_{-\vk}(t_1) \dot{\Psi}_{\vk}(t_1)-\Omega^2_{0,k} R_{-\vk}(t_1)\,\Psi_{\vk}(t_1)\Big] \nonumber\\
 &- & g^2 \int^t_0 dt_1 \int^t_0 dt_2\,R_{-\vk}(t_1) \Psi_{\vk}(t_2) \Big( G^>_k(t_1-t_2)-G^<_k(t_1-t_2)\Big)\Theta(t_1-t_2) \nonumber\\ &-& \frac{1}{2} \int^t_0 dt_1 \int^t_0 dt_2 \,R_{-\vk}(t_1) R_{\vk}(t_2)\,
 \mathcal{N}_k(t_1-t_2)  \Bigg\}\,,\label{seffina}\eea where
 \be \mathcal{N}_k(t_1-t_2)= \frac{g^2}{2} \,  \Big( G^>_k(t_1-t_2)+G^<_k(t_1-t_2)\Big) \label{noise} \ee
 and in the last term in (\ref{seffina}) we symmetrized in $t_1,t_2$ using the identity (\ref{iden}).
 In ref.\cite{qbmboy} the non-equilibrium effective action was shown to be equivalent to a stochastic description with noise whose correlation function is completely determined by the kernel (\ref{noise}).

 The path integrals over $\Psi,R$ is gaussian and can be carried out by standard methods: write
 \be R_{\vk}(t_1) = R^c_{\vk}(t_1) + r_{\vk}(t_1) ~~;~~ \Psi_{\vk}(t_1) = \Psi^c_{\vk}(t_1) + \xi_{\vk}(t_1) \,.\label{gausdef}\ee  Where   $R^c, \Psi^c$ are classical paths obeying the boundary conditions
 \bea  \Psi^c_{\vk}(0) & \equiv & \Psi_{\vk,i} = \frac{1}{2}\big(\phi_{\vk,i} + \phi^{\,'}_{\vk,i}\big)~~;~~ \Psi^c_{\vk}(t) \equiv \Psi_{\vk,f}=\frac{1}{2}\big(\phi_{\vk,f} + \phi^{\,'}_{\vk,f}\big) \label{psicbc}\\
 R^c_{\vk}(0) & \equiv & R_{\vk,i} = \big(\phi_{\vk,i} - \phi^{\,'}_{\vk,i}\big)~~~;~~~ R^c_{\vk}(t) \equiv
  R_{\vk,f} = \big(\phi_{\vk,f} - \phi^{\,'}_{\vk,f}\big) \,,\label{Rbcc} \eea   and $r(t),\xi(t)$ are the fluctuations around the classical paths and obey
 \be r_{\vk}(0) = r_{\vk}(t)=0 ~~;~~ \xi_{\vk}(0)= \xi_{\vk}(t)=0 \,. \label{rchibc}\ee $R^c,\Psi^c$ are chosen so that there is no linear term in $r_{\vk}(t_1);\xi_{\vk}(t_1)$ in the effective action, leading to the following coupled equations of motion,
 \be \ddot{\Psi}^c_{\vk}(t_1) + \Omega^2_{0,k}\Psi^c_{\vk}(t_1)+ \int^{t_1}_0 \Sigma_k(t_1-t_2) \Psi^c_{\vk}(t_2)\,dt_2 = \eta_{\vk}(t_1) \label{psiddot}\ee
 and
 \be \ddot{R}^c_{\vk}(t_1) + \Omega^2_{0,k}R^c_{\vk}(t_1)+ \int^{t}_{t_1} \Sigma_k(t_2-t_1) R^c_{\vk}(t_2)\,dt_2 = 0 \,,\label{ddotR}\ee  where
 \bea \Sigma_k(t_1-t_2) & = &  -i g^2 \Big[G^>_k(t_1-t_2)-G^<_k(t_1-t_2) \Big] \,,\label{sigma}\\
 \eta_{\vk}( t_1) & = & i   \int^t_0 \mathcal{N}_k(t_1-t_2)\,R^c_{\vk}(t_2)\,dt_2 \,.\label{eta}\eea  These equations of motion are very similar to those obtained in ref.\cite{grabert} for quantum brownian motion, they  must be solved with the boundary conditions (\ref{psicbc},\ref{Rbcc}).

  \subsection{Solutions of the equations of motion:}

  In order to obtain $S_{eff}$ we must now find the solutions to the equations of motion for $\Psi^c,R^c$.

The equation of motion (\ref{ddotR}) can be written in a form similar to (\ref{psiddot}) by introducing
\be \mathcal{Z}_{\vk}(\tau) = R^c_{\vk}(t_1) ~~;~~\tau = t-t_1 \,,\label{zetav}\ee leading to
\be \frac{d^2}{d\tau^2} \mathcal{Z}_{\vk}(\tau)+\Omega^2_{0,k}\mathcal{Z}_{\vk}(\tau)+\int^{\tau}_0 \Sigma_k(\tau-\tau')  \mathcal{Z}_{\vk}(\tau')\,d\tau' = 0 ~~;~~ \tau' = t-t_2\,.\label{eomZ}\ee

With the spectral representations of the correlation functions (\ref{specrep},\ref{rhogl}) we find

\bea  \Sigma_{k}(t-t')  & = &  -i g^2 ~ \int \frac{dk_0}{(2\pi)}\,\rho(k_0,k) \, e^{-ik_0(t-t')} \,, \label{isig} \\
\mathcal{N}_k(t-t') & = & \frac{g^2}{2}~  \int \frac{dk_0}{(2\pi)}\,\rho(k_0,k) \,\mathrm{coth\Big[ \frac{\beta k_0}{2}\Big]} e^{-ik_0(t-t')} \,, \label{noiset}\eea therefore the self energy $i\Sigma$ and $\mathcal{N}$ kernels obey the generalized fluctuation dissipation relation as shown in ref.\cite{qbmboy}.

The solutions of the   (\ref{psiddot},\ref{zetav}) are obtained by a Laplace transform.   Defining the Laplace transforms

\begin{eqnarray}
\widetilde{\Psi}^c_{\vk}(s)  \equiv \int^{\infty}_0 dt
e^{-st}\Psi^c_{\vk}(t)\label{laplapsi}\\
\widetilde{\eta}_{\vk}(s)  \equiv \int^{\infty}_0 dt
e^{-st}\eta_{\vk}(t)\label{laplaxi} \\
\widetilde{\mathcal{Z}}_{\vk}(s)\equiv \int^{\infty}_0 dt
e^{-st}\mathcal{Z}_{\vk}(t)\label{laplaZ}
\end{eqnarray}
along with
\be \widetilde{\Sigma}(k,s)\equiv \int^{\infty}_0 dt e^{-st}\Sigma_k(t)=
-\frac{g^2}{2\pi} \int^{\infty}_{-\infty}
\frac{\rho(k_0,k)}{k_0-is}~dk_0\,,
\label{laplasig}
\ee we find
\begin{equation}\label{solutionlap}
\widetilde{\Psi}^c_{\vec k}(s)=\frac{\dot{\Psi}^c_{\vk}(0)+s\Psi_{\vk,i}+\widetilde{\eta}_{\vec
k}(s)}{s^2+\Omega^2_{0,k}+\widetilde{\Sigma}(k,s)}\,,
\end{equation} and

\begin{equation}\label{solutionZ}
\widetilde{\mathcal{Z}}_{\vec k}(s)=\frac{\dot{\mathcal{Z}}_{\vk}(0)+s\mathcal{Z}_{\vk}(0)}{s^2+\Omega^2_{0,k}+\widetilde{\Sigma}(k,s)}\,.
\end{equation}

These solutions can be written more succinctly in real time in terms of the function $\mathcal{G}_k(t)$ that obeys the following equation of motion and initial conditions

\begin{equation}
\ddot{ \mg}_{k}(t_1)+\Omega^2_{0,k}\,{\mgk}(t_1)+\int_0^{t_1} dt_2 ~
\Sigma_{k}(t_1-t_2) \mgk(t_2)=0~~;~~  \mgk(t_1=0)= 0; ~~ \dot{\mg}_{
k}(t_1=0)=1 \,,\label{functionf}
\end{equation}

\noindent whose Laplace transform is given by

\begin{equation}\label{laplaf}
\widetilde{\mg}_k(s) =
\frac{1}{s^2+\Omega^2_{0,k}+\widetilde{\Sigma}(k,s)}\,.
\end{equation}
The differential equation along with the initial conditions (\ref{functionf}) imply that
\be \ddot{\mg}_k(t_1)|_{t_1=0} =0 \,. \label{secder}\ee
The function $\mgk(t_1)$ is obtained by carrying out the inverse Laplace transform,
 \be \mgk(t_1) = \int_{C}\frac{ds}{2\pi i} \frac{e^{st_1}
}{s^2+\Omega^2_{0,k}+\widetilde{\Sigma}(k,s)}\,, \label{invlap}\ee where the contour $\mathcal{C}$ is parallel to the imaginary axis in the complex s-plane to the right of all the singularities of $\widetilde{\mg}_k(s)$. Once we obtain $\mgk(t_1)$, the solutions of the
equations of motion  (\ref{psiddot}, \ref{eomZ}) are given by
\begin{equation}
\Psi^c_{\vk}(t_1) = \Psi_{\vk,i}~ \dot{\mg}_k(t_1) +
 \dot{\Psi}^c_{\vk}(0)~ \mgk(t_1)+ \int^{t_1}_0
\mgk(t_1-t_2)~\eta_{\vk}(t_2) dt_2 \,,\label{inhosolution}
\end{equation} and
\be \mathcal{Z}_{\vk}(\tau) = \mathcal{Z}_{\vk}(0)~ \dot{\mg}_k(\tau) +
 \dot{\mathcal{Z}}_{\vk}(0)~\mgk(\tau)\,. \label{soluzeta} \ee The coefficients are determined by the boundary conditions (\ref{psicbc},\ref{Rbcc}).  Using the relation (\ref{zetav}) the boundary condition (\ref{Rbcc}) yields
 \be R^c_{\vk}(t_1) =  R_{\vk,i}\,\frac{\mgk(t-t_1)}{\mgk(t)} +  R_{\vk,f}\Bigg[\dot{\mg}_{\vk}(t-t_1)-\frac{\dot{\mg}_{\vk}(t)}{\mgk(t)}\,\mgk(t-t_1)  \Bigg]\,, \label{Rcoft} \ee  and
 \be \Psi^c_{\vk}(t_1) = \Psi_{\vk,i}\Bigg[\dot{\mg}_{\vk}(t_1) -\frac{\dot{\mg}_{\vk}(t)}{{\mg}_{\vk}(t)}\, \mgk(t_1)  \Bigg]+ \Psi_{\vk,f} \, \frac{\mgk(t_1)}{\mgk(t)}+ \Bigg[ \Psi_{\vk;\eta}(t_1)-  \Psi_{\vk;\eta}(t)\,\frac{\mgk(t_1)}{\mgk(t)} \Bigg]\,,\label{psicoft} \ee where
\be  \Psi_{\vk;\eta}(t_1) = ig^2 \int^t_0 dt_2 \int^{t_1}_0 dt'\,\mgk(t_1-t') \mathcal{N}_k(t'-t_2)\,R^c_{\vk}(t_2) \,, \label{psieta} \ee in this expression $R^c_{\vk}(t_2)$ is the solution (\ref{Rcoft}).

These solutions imply a \emph{non-perturbative} Dyson resummation of one-loop self energy diagrams that yield $\mgk(t_1)$.

\vspace{1mm}

\subsection{Time evolution:}

It remains to obtain the explicit form of $\mgk(t_1)$. The function (\ref{laplaf}) generally features (complex) poles with negative real part for stability, and multiparticle cuts along the imaginary axis, hence   the contour $\mathcal{C}$ runs parallel to and to the right of the imaginary axis, namely $s= i\omega + \epsilon$, with $-\infty \leq \omega \leq \infty$, $\epsilon \rightarrow 0^+$.
 Therefore
\be \mgk(t_1) = -\int \frac{d\omega}{2\pi } \frac{e^{i\omega\,t_1}
}{\Big[(\omega-i\epsilon)^2-\Omega^2_{0,k}- {\Sigma}_R(\omega,k) -i\Sigma_I(\omega,k)\Big]} \label{goft2} \ee which is recognized as the Fourier transform of the \emph{retarded} propagator with the Dyson resummation of self-energy contributions, with
\bea \Sigma_R(\omega,k) & = &  -\frac{g^2}{2\pi}\, \int dp_0 \, \mathcal{P} \Big[\frac{\rho(p_0,k)}{p_0 - \omega} \Big] \label{sigmaR}\\
\Sigma_I(\omega,k) & = & \frac{g^2}{2} \, \rho(\omega,k) \label{sigmaI}\,, \eea where $\mathcal{P}[\cdots]$ is the principal part and  $\Sigma_R(\omega,k), \Sigma_I(\omega,k)$ are even and odd functions of $\omega$ respectively as a consequence of the property (\ref{props}).

\textbf{Renormalization:} In a renormalizable theory the real part of the self-energy is twice subtracted, and the subtractions are absorbed into mass and wave function renormalizations, therefore we write
\be \Sigma_R(\omega,k) = \Sigma_R(\Omega_k,k)+ (\omega^2 - \Omega^2_k) \,\Sigma'_R(\Omega_k,k)+ (\omega^2 - \Omega^2_k)^2 \,\widetilde{\Sigma}_R(\omega,k)\,,  \label{subsig}\ee where $\Omega_k = \sqrt{k^2+m^2_r}$ is the renormalized frequency and $m_r$ the renormalized mass. The (``on-shell'') renormalization conditions are
\be \Omega^2_{0,k}+ \Sigma_R(\Omega_k,k) = \Omega^2_k ~~;~~ Z^{-1} = \Big[1- \Sigma'_R(\Omega_k,k)\Big]\,,  \label{renconds}\ee where $Z$ is the (``on-shell'') wave function renormalization constant, and
\be Z^{-1} = 1+ g^2 \int_0^\infty \frac{dp_0}{\pi}\, \mathcal{P} \Bigg[ \frac{p_0\,\rho(p_0,k)}{(p^2_0-\Omega^2_k)^2 }\Bigg] \,,  \label{wafun}\ee where we used the property (\ref{props}).

The retarded propagator in (\ref{goft2}) now reads

\be \mathcal{G}_k(\omega,k) =  \frac{1}
{\Big[Z^{-1}\big[\omega^2-\Omega^2_{k}\big]- {\sigma}(\omega,k) -i\Sigma_I(\omega,k)-i\epsilon \omega \Big]} \,,\label{propa2}\ee where $\sigma(\omega,k) = (\omega^2 - \Omega^2_k)^2 \,\widetilde{\Sigma}_R(\omega,k)$ is finite and vanishes on the (renormalized) mass shell. This propagator features cuts along the real $\omega$-axis whenever $\rho(\omega,k)\neq 0$. The region of support of $\rho(\omega,k)$ is given by eqn. (\ref{rho2chis}).  Considering that the renormalized mass of the $\phi$ field is $m_r \ll (M_1+M_2)$ the two particle cut determined by $\rho_{2p}$ in (\ref{rho2chis}) is above the mass shell and is the \emph{only} contribution that remains as $T\rightarrow 0$.

 \begin{figure}[ht!]
\begin{center}
\includegraphics[height=3in,width=3.5in,keepaspectratio=true]{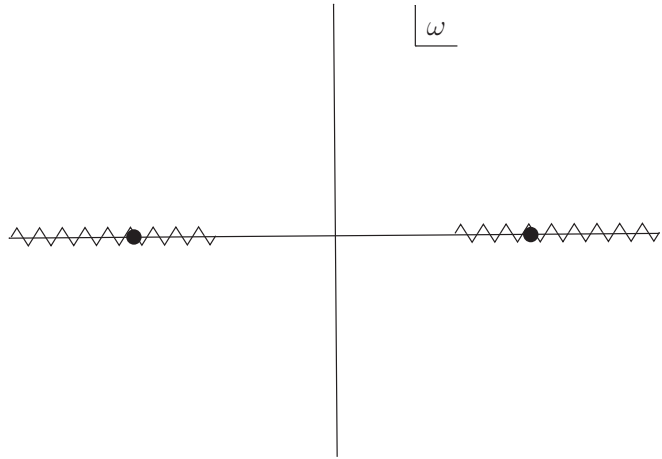}
\caption{ Analytic structure of the propagator in the resonant case, on-shell ``poles'' (denoted by the dots) are embedded in the multiparticle continuum shown with zig-zag lines.  }
\label{fig:respropagator}
\end{center}
\end{figure}

 \begin{figure}[ht!]
\begin{center}
\includegraphics[height=3in,width=3.5in,keepaspectratio=true]{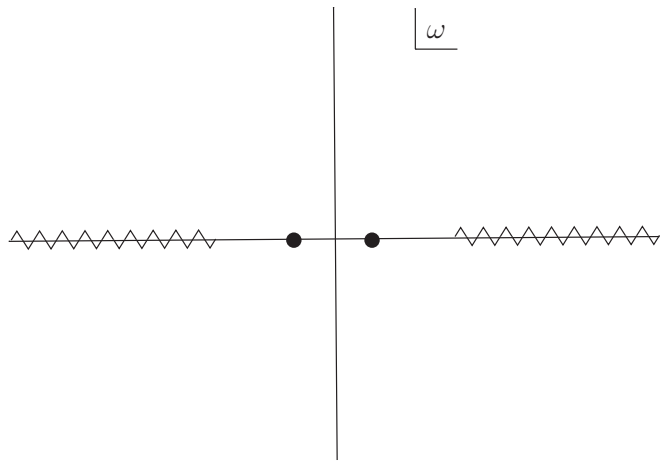}
\caption{ Analytic structure of the propagator for the non-resonant case,  isolated poles below multiparticle thresholds shown with zig-zag lines. The dots show the position of the single particle poles.  }
\label{fig:nonrespropagator}
\end{center}
\end{figure}

If $(M_1-M_2)^2 > m^2_r$ the finite temperature contribution $\rho_d$ to the spectral density (\ref{rho2chis}) has support on the $\phi$ mass shell at $Q^2 = m^2_r$, we refer to this as the \emph{resonant case}. In this case the $\phi$ on-shell ``pole'' is embedded in the continuum moving off the physical onto a second or higher Riemann sheet. On the other hand either at $T=0$ or if $(M_1-M_2)^2 < m^2_r$ the on-shell pole at $Q^2 = m^2_r$ is isolated and below the multiparticle thresholds, we refer to this as the \emph{non-resonant} case. The analytic structure of the propagator is displayed in figs.(\ref{fig:respropagator},\ref{fig:nonrespropagator})   for the resonant and non-resonant cases respectively (these figures do not display the Landau damping cut for $Q^2<0$ as it is not relevant for the discussion).

 \subsubsection{Resonant case:}

 In the resonant case, in perturbation theory the propagator in (\ref{propa2}) features \emph{complex } poles near $\omega = \pm \Omega_k$,  near these poles it can be approximated by a (narrow)  Breit-Wigner form, \be \mgk(\omega,k)= \frac{Z}{2\omega^\pm_p} \frac{1}{\Big[\omega-\omega^\pm_p-i\frac{\Gamma_k}{2}\Big]} ~~;~~\omega^\pm_p = \pm \Omega_k \,, \label{BWprop}\ee
where the width is given by
\be  {\Gamma_k}   =   Z\,\frac{\Sigma_I(\Omega_k,k)}{{\Omega_k}} = \frac{g^2\,Z}{2\Omega_k}\,\rho(\Omega_k,k)  \,.\label{width}\ee Since $Z \simeq 1 + \mathcal{O}(g^2)$ we can set $Z=1$ in (\ref{width}) to leading order in $g$. In the narrow width approximation, the complex pole in the  Breit-Wigner propagator dominates the long time limit and yields
\be \mgk(t_1) = Z\,  e^{-\frac{\Gamma_k}{2}\,t_1}\,  \frac{\sin(\Omega_k\,t_1)}{\Omega_k}\,, \label{resgoft}\ee where contributions from the continuum background are perturbatively small and  subleading in the long time limit.

\subsubsection{Non-resonant case:} In this case the isolated poles below the multiparticle thresholds dominate the dynamics at long time. The contribution at long time from the multiparticle continuum is dominated by the behavior of the $\rho(\omega,k)$ near the threshold, yielding an inverse power law decay, thus the long time behavior in this case is given by
\be \mgk(t_1) = Z\,   \frac{\sin(\Omega_k\,t_1)}{\Omega_k}+ \mathcal{F} \, \frac{\sin(\omega_{th}t_1)}{\omega_{th}\big(\omega_{th}t_1\big)^\alpha} \label{nonresgoft}\ee where $\mathcal{F}$ is a dimensionless  perturbative coefficient ($\propto  g^2$) that depends on the spectral density, $\omega_{th}$ is the threshold frequency and $\alpha$ is determined by the behavior of the spectral density near threshold\cite{boylan}. For example at $T=0$ in the case under consideration $ \omega_{th} = \sqrt{k^2+(M_1+M_2)^2}$ and $\alpha = 3/2$. With $\omega_{th} \gg \Omega_k$ the contribution from the continuum can be safely neglected for $\omega_{th} t_1  \gtrsim 1$.

We are primarily interested in obtaining the reduced density matrix in the long time limit, well after all the transient dynamics associated with the instantaneous ``switching-on'' of the coupling between the fields $\phi$ and $\psi,\chi$ has subsided. In this long-time limit we can neglect the contribution from the multiparticle (background) continuum both in the resonant and non-resonant cases. In summary, in the long time limit
\bea \mgk(t_1)   & = &  Z\,  e^{-\frac{\Gamma_k}{2}\,t_1}\,  \frac{\sin(\Omega_k\,t_1)}{\Omega_k} ~~;~~\mathrm{resonant} \label{resogt}\\
\mgk(t_1)   & = &  Z\,  \frac{\sin(\Omega_k\,t_1)}{\Omega_k} ~~;~~\mathrm{non-resonant} \,, \label{nonresogt} \eea  Therefore we can obtain the long time limit in the non-resonant case by setting $\Gamma_k \rightarrow 0$ in the resonant case.

We now have all the necessary ingredients to obtain the effective action and carry out the (functional) integral over the initial values. In the solutions (\ref{Rcoft},\ref{psicoft}) we   take the long time limit for $t$ with the results (\ref{resogt},\ref{nonresogt}). Furthermore, in the   integrals up to time $t$ in the quadratic term in $R^c$ in $S_{eff}$ (\ref{seffin}) and in $\Psi_{\vk,\eta}(t)$ in (\ref{psicoft},\ref{psieta}) we carry out these integrals taking $t\rightarrow \infty$, yielding the following results:
\be \dot{\Psi}_{\vk,\eta}(0)= 0 ~~;~~ \dot{\Psi}_{\vk,\eta}(\infty)= i R_{\vk,f} \,K_F \label{dotPsieta}\ee
\be \Psi_{\vk,\eta}(\infty) = i R_{\vk,i}\,\eta_I + i R_{\vk,f}\,\eta_F \,,\label{Psietaasy} \ee
\be \int^\infty_0 dt_1 \int^\infty_0 dt_2 R^c_{-\vk}(t_1)\,R^c_{\vk}(t_2) \mathcal{N}_{k}(t_1-t_2) = R^2_{\vk,i}\,J_1 + R^2_{\vk,f}\,J_2+2 R_{\vk,i}R_{\vk,f}\,J_3\,, \label{quadR}\ee
\be \dot{\Psi}^c_{\vk}(0) = -   \Psi_{\vk,i}\Bigg[ \frac{\dot{\mg}_{\vk}(t)}{{\mg}_{\vk}(t)}\, \Bigg]+   \frac{\Psi_{\vk,f}}{\mgk(t)}- i\Bigg[    \frac{   R_{\vk,i}\,\eta_I + R_{\vk,f}\,\eta_F }{\mgk(t)} \Bigg]\,,\label{dotpsicofti}\ee
 \be \dot{\Psi}^c_{\vk}(t) = \Psi_{\vk,i}\Bigg[\ddot{\mg}_{\vk}(t) -\frac{\dot{\mg}_{\vk}(t)}{{\mg}_{\vk}(t)}\, \dot{\mg}_k(t)  \Bigg]+ \Psi_{\vk,f} \, \frac{\dot{\mg}_k(t)}{\mgk(t)}+ i\,\Bigg[    R_{\vk,f} \,K_F-   \big( R_{\vk,i}\,\eta_I +  R_{\vk,f}\,\eta_F \big) \,\frac{\dot{\mg}_k(t)}{\mgk(t)} \Bigg]\,,\label{dotpsicoftf} \ee
 where we used the result (\ref{secder}) in (\ref{dotpsicofti}) and $K_F,\eta_I,\eta_F,J_{1,2,3}$ are given explicitly in appendix (\ref{sec:coeffs}) for the resonant case with $\gamma=\Gamma_k/2$. The non-resonant case is obtained from these expressions by setting $\gamma=0$.

 In terms of the solutions of the equations of motion with the proper boundary conditions found above  we find
\be iS_{eff}=  \sum_{\vk}\Bigg\{i\,\Big[R_{-\vk,f} \dot{\Psi}^c_{\vk}(t) - R_{-\vk,i} \dot{\Psi}^c_{\vk}(0) \Big] + \frac{1}{2}\int^t_0 dt_1 \int^t_0 dt_2 R^c_{-\vk}(t_1)\,R^c_{\vk}(t_2) \mathcal{N}_{k}(t_1-t_2) \Bigg\} + i\widetilde{\mathcal{S}}(t)\,. \label{seffin}\ee The term $i\widetilde{\mathcal{S}}(t)$ arises from the fluctuations $r_{\vk},\xi_{\vk}$ with   boundary conditions (\ref{rchibc}), it does not depend on the boundary values of the fields $\Psi_{\vk,i,f},R_{\vk,i,f}$ and only depends on time. Its contribution to the path integral (propagating kernel) is
\be e^{i\widetilde{\mathcal{S}}(t)}\equiv \widetilde{N}(t)\,, \label{normat}\ee this is an overall time dependent normalization factor. We do \emph{not} need to calculate this factor because it is completely determined by the identity (\ref{normafina}), a consequence of unitary time evolution of the full density matrix. Once we obtain $S_{eff}$ for $\widetilde{\mathcal{S}}(t)=0$ the normalization is fixed by the identity (\ref{normafina}). In the following discussion we refer to $S_{eff}$ as the effective action with $\widetilde{\mathcal{S}}(t)=0$, we will account for this normalization factor at the end of the calculation.

We emphasize that the effective action (\ref{seffin}) results from a \emph{non-perturbative} Dyson resummation of one-loop diagrams, therefore yields the time evolution of the reduced density matrix beyond the perturbative analysis of section (\ref{sec:modelpert}).

\section{Von-Neumann Entropy}

With  the final form for $\Psi^c,R^c$ given above,
the effective action (\ref{seffin}) for $\widetilde{\mathcal{S}}=0$ can be obtained straightforwardly by replacing (\ref{quadR},\ref{dotpsicofti},\ref{dotpsicoftf}) in (\ref{seffin}) (for $\widetilde{\mathcal{S}}=0$), it depends explicitly on $\Psi_{\vk,i},\Psi_{\vk,f};R_{\vk,i},R_{\vk,f}$ and time $t$. The reduced density matrix (\ref{rhorfin}) with the initial density matrix  (\ref{rhoinift}) is given by
 \be \rho^{r}(\phi_f,\phi'_f;t) = \widetilde{N}\,\int D\Psi_i   DR_i  \,  \, e^{iS_{eff}[\Psi_{\vk,i},\Psi_{\vk,f};R_{\vk,i},R_{\vk,f};t]} ~~ e^{-\sum_{\vk}\Omega_{0k} \Big[\Psi_{\vk,i}\Psi_{-\vk,i}+\frac{1}{4}\,R_{\vk,i}R_{-\vk,i}\Big]} \,, \label{rhorfinal} \ee where we used   $D \phi_i D\phi'_i = D\Psi_i D R_i$  along with  (\ref{psibc},\ref{Rbc}).  The overall normalization factor $\widetilde{N}$ is fixed by the condition (\ref{normafina}). The functional integrals over $\Psi_i,R_i$ are Gaussian and carried out straightforwardly, the general final form of the reduced density matrix is found to be
 \be    \rho^{r}(\Psi_f,R_f;t) = N(t)\, e^{-\sum_{\vk}\Big[A_k(t)\, \Psi_{\vk,f}\Psi_{-\vk,f}+ B_k(t)\, R_{\vk,f}R_{-\vk,f}+i\,C_k (t) \,\Psi_{\vk,f}R_{-\vk,f}\Big]} \label{rhofinalis}\ee The coefficients $A_k(t),B_k(t),C_k(t)$ are all real   functions of  the various coefficients $K_F, \eta_I,\cdots $ and depend explicitly on time $t$. They will be obtained in the long time limit for the non-resonant and resonant cases separately below.

 In terms of $\phi_f, \phi'_f$ the reduced density matrix reads
 \be  \rho^{r}(\phi_f,\phi'_f;t) = N(t)\,\Pi_{\vk}~ e^{-\Big[\Big(\frac{A_k}{4}+B_k\Big)\,\Big( (\phi_{\vk,f})^2+ (\phi'_{\vk,f})^2 \Big)-2\phi_{\vk,f}\phi'_{-\vk,f}\Big(B_k-\frac{A_k}{4}\Big)\Big]}~~
 e^{-iC_k \Big( (\phi_{\vk,f})^2- (\phi'_{\vk,f})^2\Big)}\,, \label{rhorfifip}\ee where $(\phi_{\vk,f})^2 \equiv \phi_{\vk,f}\phi_{-\vk,f}, \mathrm{etc}$, as shortand notation.

  Although there are several alternative definitions of entropy\cite{qinfo}, we focus on obtaining the Von-Neumann entropy because it has a natural thermodynamic interpretation and allows us to study the possibility of thermalization.

  The Von-Neumann entropy is obtained from the  eigenvalues $p_n$ of the density matrix
 \be \int D\phi'_f \rho^r(\phi_f,\phi'_f) \Phi_n[\phi'_f] = p_n \Phi_n[\phi_f] \,, \label{pns}\ee namely
 \be S = -\sum_{n} p_n \,\ln[p_n] \,.\label{VNS} \ee  We note that the   phase $e^{-iC_k \Big( (\phi_{\vk,f})^2- (\phi'_{\vk,f})^2\Big)}$ in (\ref{rhorfifip}) does \emph{not} contribute to the eigenvalue equation:   this phase is absorbed into a redefinition of the wave functionals, namely, $\Phi_n[\phi_f] \rightarrow e^{iC_k (\phi_{f})^2}\, \Phi_n[\phi_f]$, therefore   we     set $C_k=0$ in (\ref{rhorfifip}). In ref.\cite{sorkin} the Von-Neumann entropy for generic Gaussian density matrices has been obtained, using the results from this reference along with those  from refs.\cite{GR,srednicki,neu} we find (up to an overall normalization)
 \be \Phi_n[\phi_f] = H_n(\sqrt{\omega_k}\,\phi_f)\,e^{-\omega_k (\phi_f)^2/2} ~~;~~ \omega_k = 2\sqrt{B_k A_k} \label{wavefunc}\ee where $H_n$ are Hermite polynomials, and imposing the normalization condition (\ref{normafina}), which results in $\sum_n p_n =1$, we find
 \be p_n = \frac{2}{\alpha_k+1} \Bigg[ \frac{\alpha_k-1}{\alpha_k+1} \Bigg]^n ~~;~~ \alpha_k= \sqrt{\frac{4B_k}{A_k}}\,. \label{probs}\ee This result has a more illuminating interpretation in terms of a thermal density matrix: introduce momentum and time-dependent frequency $\omega_k(t)$ and effective ``temperature'' $ T_k(t)$ via the following relations valid for $4B_k \geq  A_k$ (this inequality will be confirmed explicitly below)
 \be B_k(t)+\frac{A_k(t)}{4} \equiv \frac{\omega_k(t)}{2} \,\coth\Big[\frac{\omega_k(t)}{T_k(t)}\Big]   ~~;~~  B_k(t)-\frac{A_k(t)}{4} \equiv \frac{\omega_k(t)}{2\,\sinh\Big[\frac{\omega_k(t)}{T_k(t)}\Big]}\label{omegaTNR}\ee where we have now exhibited the $t$ dependence of the coefficients explicitly. It follows from (\ref{omegaTNR}) that
 \be \omega_k(t) = 2\sqrt{B_k(t) A_k(t)} ~~;~~ e^{-\frac{\omega_k(t)}{T_k(t)}} = \frac{\alpha_k-1}{\alpha_k+1}\,, \label{omes}\ee with $\alpha_k$ given by eqn. (\ref{probs}).
 Replacing the definitions (\ref{omegaTNR}) in the reduced density matrix (\ref{rhorfifip}) and setting $C_k =0$ according to the discussion above, we find
  \be  \rho^{r}(\phi_f,\phi'_f;t) = N(t)\,\Pi_{\vk}~ e^{- \frac{\omega_k(t) }{2\,\sinh\big[\frac{\omega_k(t)}{T_k(t)}\big]}\, \Big[\cosh\big[\frac{\omega_k(t)}{T_k(t)}\big]\,\Big( (\phi_{\vk,f})^2+ (\phi'_{\vk,f})^2 \Big)-2\phi_{\vk,f}\phi'_{-\vk,f} \Big]}\,, \label{rhorT}\ee This is a thermal density matrix for a free Gaussian field of frequency $\omega_k(t)$ at an effective temperature $T_k(t)$ for each $\vk$\cite{feynman}. Fixing the overall normalization from the condition (\ref{normafina}), the eigenvalues of this density matrix are the thermal probabilities,
  \be p_n= \big[1-e^{-\frac{\omega_k(t)}{T_k(t)}}\big]\, \Big(e^{-\frac{\omega_k(t)}{T_k(t)}} \Big)^n \,\label{thermalpn}\ee which coincide with (\ref{probs}) via the definitions (\ref{omes}). Since the total reduced density matrix factorizes into a product for each independent $\vk$, the total Von-Neumann entropy is given by
  \be S = -\sum_{\vk} \Big[ \ln(1-\zeta_k(t))+ \frac{\zeta_k(t)\,\ln(\zeta_k(t))}{(1-\zeta_k(t))}\Big]~~;~~\zeta_k(t) = e^{-\frac{\omega_k(t)}{T_k(t)}} \,. \label{totentropy}\ee This is the same expression obtained in refs.\cite{srednicki,callan} for the geometric entropy with the variable $\zeta$ given by a different function of parameters.

  Before we proceed to obtain $S$ for the non-resonant and resonant cases, we comment on several noteworthy aspects of the analysis above.

  \vspace{1mm}

  \textbf{a:)} It is clear from the result (\ref{rhorfifip}) that if $4 B_k = A_k$ the reduced density matrix is of the form $\Phi[\phi_f]\,\Phi^*[\phi'_f]$, namely it describes a \emph{pure state}, however if $B_k + A_k/4 = A_k/2 \neq \Omega_k/2$  it is a two-mode squeezed state\cite{zubairy}. A non-vanishing entropy arises from the term (in the exponent) that is linear in $\phi_f$ and linear in $\phi'_f$ (the term proportional to $B_k-A_k/4$ in eqn. (\ref{rhorfifip})). When the field is expanded in creation and annihilation operators (\ref{friex}), these \emph{linear} terms are associated with \emph{single particle states}, unlike the quadratic terms in $\phi_f,\phi'_f$ which are associated with \emph{pairs}. It is this (linear) term the one associated with the purity of the density matrix and the entropy, thus establishing a direct correspondence with the perturbative analysis in section (\ref{sec:modelpert}) which concluded that the mixed nature of the reduced density matrix  at second order (\ref{rhosecor}) is encoded in the contribution from single particle states (see discussion after eqn. (\ref{alfastate})).

  \vspace{1mm}

  \textbf{b:)} A probability interpretation of the eigenvalues of the density matrix is only available provided $4B_k \geq A_k$. It is not \emph{a priori} evident that this condition is fulfilled, however, it will be shown below to be fulfilled explicitly
  both in the non-resonant and resonant cases   in the asymptotic long time limit.

  \vspace{1mm}

  \textbf{c:)} At this stage, the parameters $T_k(t)$ describe an \emph{effective} temperature because of the similarity of the reduced density matrix to a thermal one  even when the $\phi,\chi$ fields are \emph{initially in their ground state} which corresponds to $T=0$ when the spectral density is given by (\ref{zeroTrho2chi}), namely the non-resonant case. This similarity suggests that the reduced density matrix describes an \emph{incipient} thermalization albeit with a \emph{non-equilibrium  temperature} for each individual mode.

\vspace{1mm}

\subsection{Non-resonant case:}

For the non-resonant case we set $\Gamma_k =0$ since the spectral density does not have support on the (renormalized) mass shell at $k_0 = \Omega_k$. The coefficients are obtained from the results of appendix B by setting $\gamma=0$. As discussed above we neglect the coefficient $C_k$ since it does not contribute to the eigenvalues of the reduced density matrix. After straightforward algebra we find:
\be A_k = \frac{1}{Z^2}\,\Big[\frac{\Omega_{0k}\,\Omega^2_k}{\Omega^2_{0k}\mathcal{S}^2+\Omega^2_k \mathcal{C}^2}\Big]-2\,\Omega^2_k \,F_k ~~;~~ \mathcal{S} = \sin(\Omega_k t)~~;~~\mathcal{C} = \cos(\Omega_k t) \label{ANR}\ee
\be B_k = \frac{Z^2}{4}\,\Big[\frac{\Omega_{0k}\,\Omega^2_k}{\Omega^2_{0k}\mathcal{S}^2+\Omega^2_k \mathcal{C}^2}\Big]+ \frac{K_F}{2}  \label{BNR}\ee where $F_k$ and $K_F$ are given by (\ref{Ffina},\ref{KF}) for $\gamma=0$ respectively,  $Z$ is given by (\ref{wafun}) and $\Omega^2_{0k}, \Omega^2_k$ are related by mass renormalization (\ref{renconds}). Writing $Z = 1+ g^2\,z_1+\cdots $ where $z_1$ can be read off (\ref{wafun}), using that $\Omega^2_{0k}-\Omega^2_k \propto g^2$ from (\ref{renconds}),   and gathering terms up to order $g^2$ we find
\be \frac{4B_k}{A_k} = 1 +  \Big(  4 \, g^2   z_1+ \frac{2  K_F}{\Omega_k} + 2   \Omega_k F_k \Big)+\cdots \,.\label{ratNR} \ee where the dots stand for terms of higher order in $g$.

Therefore,  up to order $\mathcal{O}(g^2)$ we   find for the non-resonant case
\be \frac{4B_k}{A_k} = 1+ \frac{2g^2}{\Omega_k} \int^\infty_0 \frac{dp_0}{\pi} \frac{\rho(p_0,k)}{(p^2_0-\Omega^2_k)^2}\Bigg[(p_0-\Omega_k)^2+ 2 ~ \frac{(p^2_0+\Omega^2_k)}{e^{p_0/T}-1} \Bigg]   > 1 \label{ratioNR}\ee where the equality is a consequence of   $\rho(p_0 >0,k) >0$. From the relations (\ref{probs},\ref{omes},\ref{totentropy})) we find
\be \zeta_k(t) = e^{-\frac{\omega_k(t)}{T_k(t)}} = \frac{g^2}{2\Omega_k} \int^\infty_0 \frac{dp_0}{\pi} \frac{\rho(p_0,k)}{(p^2_0-\Omega^2_k)^2}\Bigg[(p_0-\Omega_k)^2+ 2 ~ \frac{(p^2_0+\Omega^2_k)}{e^{p_0/T}-1} \Bigg] +\cdots \label{zetaNR}\ee Note that the finite temperature correction is manifestly positive thereby \emph{increasing} $\zeta_k(t)$ and  the entropy.

To leading order in the coupling, we find at $T=0$ the \emph{entanglement entropy density}
\be \frac{S}{V} = - {{g^2}} \,\int \frac{d^3k}{(2\pi)^3\,\Omega_k}\,  \int^\infty_0 \frac{dp_0}{2\pi} \frac{\rho(p_0,k)}{(p_0+\Omega_k)^2}~\ln\Bigg[\frac{g^2}{\Omega_k} \int^\infty_0  \frac{dp_0}{2\pi } \frac{\rho(p_0,k)}{(p_0+\Omega_k)^2} \Bigg] \,.  \label{ententNR}\ee

The  result (\ref{ententNR}) is noteworthy: at $T=0$ the initial density matrix describes a pure state, corresponding to a tensor product of the ground states for the light and heavy fields, and vanishing entropy. However tracing out the heavy degrees of freedom in the time evolution leads to an asymptotic reduced density matrix that describes a
\emph{mixed state} which resembles a thermal density matrix with an effective coupling dependent temperature for each mode. This mixed state results from the \emph{entanglement} between the light and heavy fields via their interaction as exhibited in the perturbative evaluation of the reduced density matrix in section (\ref{sec:modelpert}). Therefore this entropy is identified with the \emph{entanglement entropy}, the growth of entropy is a consequence of tracing over the heavy degrees of freedom. The coupling dependence of the entanglement entropy in this case is in agreement with the perturbative arguments in section (\ref{sec:modelpert}).

We note that the effective frequency $\omega_k(t) = 2\sqrt{B_k A_k}$ is \emph{time dependent}, however the ratio $e^{-\omega_k(t)/T_k(t)}$ given by (\ref{zetaNR}) is time independent, hence   $T_k(t)$ is also time dependent. The origin of the time dependence of $\omega_k(t)$ (hence of $T_k(t)$) can be traced to the initial density matrix (\ref{rhoinift}) in terms of the bare frequency $\Omega_{0k}$. If, instead, the initial density matrix were to be given in terms of the \emph{renormalized} frequency $\Omega_k$, then from (\ref{ANR},\ref{BNR}) it is clear that $\omega_k(t)$ (and $T_k(t)$) would be \emph{time independent}. Therefore the time dependence of the effective frequency and temperature is a manifestation of the \emph{memory of the initial conditions}. We conclude that generally, the density matrix \emph{does not relax} to a stationary state and even in the asymptotic long time limit and it retains memory of the initial condition, although the Von-Neumann entanglement entropy reaches an asymptotic stationary value. Only if the initial density matrix corresponds to the renormalized state, the asymptotic long time limit leads to a time independent stationary state in the sense that not only the entanglement entropy but also both the effective frequency and temperature become time independent.

\subsection{Resonant case:} In the resonant case $\Omega_k$ is embedded in the continuum, namely above the multiparticle threshold  and $\rho(\Omega_k,k) \neq 0$. In this case, all integrals in appendix (\ref{sec:coeffs}) that yield the coeffients $K_F,\eta_I,\cdots$ are dominated by the sharp resonances at $p_0 = \pm \Omega_k$, as a result, to leading order in $g^2$ we find
\be K_F = \Omega^2_k F_k ~~;~~ J_1 = \frac{F_k}{\mgk^2(t)}   ~~;~~ J_2 = \Omega^2_k  \frac{F_k}{\mgk^2(t)} ~~;~~ J_3 = - \Omega_k \frac{\cos(\Omega_k t)}{\sin(\Omega_k t)}\, \frac{F_k}{\mgk(t)} \,.\label{rescoefs}\ee In the long time limit $\Gamma_k t \rightarrow \infty$, we neglect terms   proportional to $\mgk(t)$ that vanish exponentially, with the result that
\be A_k = \frac{1}{2 F_k}~~;~~ B_k = \frac{1}{2}\, F_k \, \Omega^2_k \label{ABRes}\ee where $F_k$ is given by eqn. (\ref{funcF}) and again neglecting the (real) coefficient $C_k$ which does not contribute to the probabilities. In the narrow width limit with $\gamma = \Gamma_k /2 \ll \Omega_k$ we find
\be F_k = \frac{g^2 \rho(\Omega_k,k)}{4\Omega^2_k\,\Gamma_k}\,\coth\Big[ \frac{\Omega_k}{2T}\Big]  = \frac{1}{2\Omega_k}\, \coth\Big[ \frac{\Omega_k}{2T}\Big] \,,\label{FofkRes} \ee where we have used eqn. (\ref{width}) setting $Z=1$ to leading order in $g^2$  to arrive at the last equality. Therefore, in the narrow resonance limit we find
\be A_k = \frac{ \Omega_k }{\coth\Big[ \frac{\Omega_k}{2T}\Big]}+ \cdots~~;~~ B_k = \frac{\Omega_k}{4}\,\coth\Big[ \frac{\Omega_k}{2T}\Big]+\cdots  \,,\label{ABresfin}\ee where the dots stand for terms of $\mathcal{O}(g^2)$, leading to
 \be B_k(t)+\frac{A_k(t)}{4} =  \frac{\Omega_k}{2}  \coth\big[\frac{\Omega_k}{T}\big] +\cdots     ~~;~~  B_k(t)-\frac{A_k(t)}{4} =  \frac{\Omega_k }{2\,\sinh\big[\frac{\Omega_k}{T}\big]} +\cdots  \label{omegaTres}\ee This is a remarkable result, the reduced density matrix describes an equilibrium state of a free field of renormalized frequency $\Omega_k$ at temperature $T$, which is the common equilibrium temperature of the heavy fields. Now the probabilities $p_n$ are the thermal probabilities
\be p_n= \big[1-e^{-\frac{\Omega_k }{T}}\big]\, \Big(e^{-\frac{\Omega_k}{T}} \Big)^n \,\,\label{thermalpnRes}\ee  and the total entropy associated with the reduced density matrix is simply the \emph{thermal entropy} of a free field at equilibrium temperature $T$  but with renormalized mass,
 \be \frac{S^r}{V} = -\int \frac{d^3k}{(2\pi)^3}  \Big[ \ln(1-\zeta_k)+ \frac{\zeta_k\,\ln(\zeta_k)}{(1-\zeta_k )}\Big]+ \cdots ~~;~~\zeta_k = e^{-\frac{\Omega_k}{T}} \,. \label{totentropyRes}\ee

 The conclusion is that in the resonant case, the light field \emph{thermalizes} with the heavy fields on a relaxation  time scale $t_{th}(k) \simeq 1/\Gamma_k$. In the weak coupling limit and when $T \ll M_1,M_2$ this relaxation time may be very long, but ultimately the reduced density matrix for the light field becomes a thermal density matrix for a weakly interacting light field.

 For $T\neq 0$ the initial density matrix (\ref{inidensmtx}) with (\ref{rhofini}) and (\ref{rhochi}) describes the free heavy fields in thermal equilibrium, therefore the entropy of the initial state is
 \be \frac{S^{in}}{V} = -\sum_{a =1,2}  \int \frac{d^3k}{(2\pi)^3}  \Big[ \ln(1-\zeta^{a}_k)+ \frac{\zeta^a_k\,\ln(\zeta^a_k)}{(1-\zeta^a_k )}\Big]~~;~~\zeta^a_k = e^{-\frac{W^a_k}{T}} \,,  \label{inientropy}\ee with ${W^a_k} = \sqrt{k^2+M^2_a} $. With $M_{1,2} \gg m_r$ it follows that $\zeta_k \gg \zeta^a_k ~~;~~ a = 1,2$  leading to the inequality
 \be S^r > S^{in}\,, \label{entropyprod}\ee implying that entropy has increased during the time evolution as a consequence of  tracing out the heavy fields.

 In obtaining the results (\ref{ABresfin}) we have consistently neglected terms of $\mathcal{O}(g^2)$ to leading order,  keeping these terms would require to also include  higher order loop corrections ($\mathcal{O}(g^4)$).

 \section{Discussion:}
 Several aspects of the results obtained above for the non-resonant and resonant cases merit discussion.

 \vspace{1mm}

 \textbf{i:)}
 The time dependence of $\omega_k(t)$ which leads to the time dependence of the effective temperature $T_k(t)$ in the non-resonant case can be shown to originate in the first term in eqn. (\ref{dotpsicoftf}) and can be understood by setting $g^2=0$, but taking $\mgk = \sin(\Omega_k t)/\Omega_k$. This corresponds to evolving an initial state which is simply the ground state of a free field with the \emph{bare} mass and bare frequency $\Omega_{0k}$ with a free field Hamiltonian of the field with the \emph{renormalized} mass and renormalized frequency $\Omega_k$. The initial density matrix does not commute with the evolution Hamiltonian and the time dependence of $\omega_k(t)$ is a consequence of ``squeezing'' the initial state by the creation-annihilation of virtual pairs, explaining the time evolution with $\cos(2\Omega_k t)$,  namely \emph{twice} the frequency $\Omega_k$ corresponding to the eigenstates of the renormalized Hamiltonian. This is further understood if $\Omega_{0k} \rightarrow \Omega_k$ in the initial state which now is an eigenstate of the free field Hamiltonian with the renormalized mass. In the non-resonant case the the first term in (\ref{dotpsicoftf}) remains in the long time limit and oscillates if $\Omega_{0k}\neq \Omega_k$ indicating that in this case the reduced density matrix retains memory of the initial condition. This memory results in the time dependence of $\omega_k(t)$ and consequently of $T_k(t)$, when $\Omega_{0k}\neq \Omega_k$, however the entanglement entropy is independent of time.

 In the resonant case the   the first term in (\ref{dotpsicoftf})
 \emph{vanishes}   in the long time limit and the reduced density matrix looses memory of the initial state on the time scale $1/\Gamma_k$.

 \vspace{1mm}

 \textbf{ii:)} For the non-resonant case at $T=0$   the entanglement entropy  is given by (\ref{ententNR}). In the super-renormalizable model discussed here the spectral density approaches a constant as $p_0 \rightarrow \infty$ and the $p_0$ integrals in the entanglement entropy are finite. However, the k-integral    \emph{diverges} with an upper momentum cutoff $\Lambda$, yielding
 \be \frac{S_{nr}}{V} \simeq g^2 \Lambda \,\ln\Big[ \frac{\Lambda^2}{g^2} \Big]\,.  \label{Snrdiv}\ee

In a renormalizable theory,  for example if the light scalar $\phi$ is Yukawa coupled to heavy fermions, the spectral density grows $\propto p^2_0$ as $p_0 \rightarrow \infty$ and the $p_0$ integrals diverge linearly with  an upper frequency-momentum cutoff $\Lambda$. In the non-resonant case, this divergence in a renormalizable theory yields an entanglement entropy $ S_{nr} \propto g^2 (L/l)^3 \ln (\tilde{M}\,l)$ with $L=V^{1/3}$ the size of the system,   $l \propto 1/\Lambda$ and $\tilde{M}$ a scale associated with the spectral density (in a renormalizable theory $g$ is dimensionless). The dependence on the coupling is a result of tracing the heavy degrees of freedom and has also been noticed within a different setting in ref.\cite{giddings}. The geometric entropy obtained in refs. \cite{callan,srednicki,cardy} from tracing out the degrees of freedom within a spatial domain is $S_{geo}\propto A/l^2$ where $A$ is the area that separates the spatial domain and $l$ a short distance cutoff, in the case of the Bekenstein-Hawking Black Hole entropy $l$ is the Planck scale.

 \vspace{1mm}

 \textbf{iii:)} In the resonant case the asymptotic long time limit yields a reduced density matrix that is thermal at temperature $T$, the common equilibrium temperature of the heavy fields. It may be argued that an effective field theory is a suitable description for $T \ll M_1, M_2$, namely for scales well below the energy scale of the heavy fields. In this case the finite temperature contribution to the spectral density will be suppressed by terms of the form $e^{-M_{1,2}/T}$ and so is the relaxation rate $\Gamma_k$ given by (\ref{width}). Nevertheless, however small $\Gamma_k$ is,  at asymptotically long time $t\gg 1/\Gamma_k$ the reduced density matrix will relax to a thermal density matrix and if $T \gg m_r$ the light quanta will have a large occupation number and with a thermal entropy much larger than the entropy of the initial state. The entropy increase, from the thermal entropy for the heavy fields, to the thermal entropy of the light fields (with much larger occupation number) is  a consequence of tracing out the heavy degrees of freedom and the concomitant loss of information.

 \section{Conclusions and further questions:}

Our study is motivated by the observation that an effective field theory describing phenomena below some energy scale  emerges,  fundamentally,   from tracing over or coarse graining the high energy degrees of freedom with a concomitant loss of information. The information loss entailed in the reduction from the full theory where the low and high energy degrees of freedom are coupled, to the effective theory describing the low energy sector  has to be manifest as an increase in the entropy. In this article we study the information loss as a consequence of tracing over high energy degrees of freedom in a model of a light scalar field coupled to two other heavy scalar fields. The initial density matrix is factorized into the ground state for the light field and a thermal ensemble at a common temperature $T$ for the heavy fields. The case of the heavy fields in their ground state is recovered in the $T\rightarrow 0$ limit.  This initial state is evolved in time with the full interacting Hamiltonian and the heavy degrees of freedom are traced out from the time evolved density matrix up to second order in the coupling, yielding a reduced density matrix for the light degrees of freedom. The time evolution of the reduced density matrix is determined by the \emph{effective action} and defines the effective field theory description of the dynamics of the light degrees of freedom.   From the reduced density matrix we obtain the Von Neumann  entropy in the asymptotic long time limit.

 We begin with a perturbative study which exhibits explicitly    the emergence of a mixed state upon tracing the heavy degrees of freedom and   the origin of the  entropy in the \emph{entanglement} between the light and heavy degrees of freedom as a consequence of their mutual interaction, namely the entropy is recognized as a the \emph{entanglement entropy}.

 We then obtain the effective action in a non-perturbative Dyson resummation of one-loop correlations of the heavy fields.   There are two important cases: \textbf{i:)} when the (renormalized) mass shell of the light field is below the multiparticle thresholds in the spectral density of the correlation functions of the heavy fields, we refer to this as the \emph{non-resonant} case, and \textbf{ii:)} when it is within the continuum (above thresholds) and the spectral density has support on the renormalized mass shell of the light field, we refer to this as the \emph{resonant} case. When the masses of the heavy fields are much larger than that of the light field, the resonant case can only occur for $T\neq 0$ when the mass difference of heavy fields is larger than the mass of the light field\cite{qbmboy}. In the non-resonant case the reduced density matrix is formally equivalent to a thermal density matrix of a free field although with a momentum, time and coupling dependent \emph{effective} temperature. In this case the entanglement entropy is given by equation (\ref{ententNR}) it depends on the coupling between light and heavy degrees of freedom and the details of the spectral density. In this non-resonant case the reduced density matrix retains \emph{memory} of the initial conditions even at asymptotically long time, and the time dependence of the effective temperature is a consequence of squeezing in the initial state. We show that the entanglement entropy in the non-resonant case features \emph{ultraviolet} divergences. In the super-renormalizable case we find up to logarithmic corrections $S \propto g^2 L^3 \Lambda$ and  for a renormalizable theory   we argue that, up to logarithmic corrections,  $S \propto g^2  (L\Lambda)^3$    with $g$ the heavy-light coupling, $L$ the linear size of the system and $\Lambda$ an ultraviolet cutoff.

In the resonant case the light field \emph{thermalizes} with the heavy fields at temperature $T$, the reduced density matrix is completely determined by the thermal density matrix of a (nearly) free field of renormalized mass at temperature T, the (common) equilibrium temperature of the heavy fields. Therefore the entanglement entropy in this case coincides with the thermal entropy. In both cases we show that the coarse graining (tracing) procedure that yields the effective field theory leads to an \emph{increase} in the entropy concomitant with the loss of information in the coarse graining procedure. In conclusion, our study demonstrates quantitatively that effective field theories, emerging from integrating out high energy degrees of freedom are characterized by an information loss. This is manifest in the entanglement entropy associated with the reduced density matrix of the low energy degrees of freedom which in the case of thermalization becomes the thermal entropy. Although we have studied a particular model, the resonant and non-resonant cases are of broader relevance as they highlight respectively the dependence of the entanglement entropy on the couplings to and spectral densities of the heavy fields, features that are, arguably, qualitatively fairly robust.

Furthermore, within the context of cosmology, this study   strongly suggests that  the entanglement entropy associated with the decoupling of heavy fields that interact with a light sector must also be included in the entropy budget.

\vspace{1mm}

\textbf{Further questions:}   several further questions   merit  exploration.

\textbf{1:}  Are there phenomenological consequences or observables associated with the entanglement entropy of the effective field theory?. In ref.\cite{lello1} it was argued that in the case of particle decay, the products are kinematically entangled and if one (or more) are not detected (``invisible'') tracing them out of the final state yields a density matrix which features an entanglement entropy. This reference suggested possible experimental probes of this entropy. Similar conclusions but within different settings were obtained in refs.\cite{seki,semenoff} for scattering experiments. However, these discussions in the literature do not directly address the issue of entanglement within the context of an  effective field theory. Entanglement between super and sub-Hubble degrees of freedom with a concomitant entanglement entropy has been discussed in ref.\cite{lelloentcos} within the context of inflationary cosmology, is there an observable cosmological consequence of this phenomenon?. Within the cosmological context, there is also the question of how does a particle species whose distribution function has frozen out of equilibrium contribute to the entropy budget.

\textbf{2:} The entanglement entropy is revealed in the reduced density matrix arising from tracing heavy degrees of freedom out of the total time evolved density matrix and becomes manifest in the in-in field theory formulation. The usual treatment of an effective field theory typically begins with writing down an effective local Lagrangian density and performing in-out (S-matrix) calculations of low energy observables. In these calculations there is no hint of  entanglement between the low and high energy degrees of freedom or its consequence, the entanglement entropy. Neither cross sections in the effective field theory nor any other observable calculated from S-matrix elements bear any relation to the entanglement entropy or any alternative quantity related to information loss. The matching between the high energy and low energy physics that is necessary to obtain the effective couplings of the effective field theory is often cast in terms of Wilson's coefficients in an operator product expansion, are these coefficients in any way related to the entanglement entropy?, perhaps with correspondence to the short distance divergences of the entanglement entropy   discussed above?.

\textbf{3:} We have argued that in a renormalizable theory, the entanglement entropy in the non-resonant case features divergences proportional to $\Lambda^3$ with $\Lambda$ an ultraviolet cutoff, in four  space-time dimensions the geometric entropy features a divergence $\propto \Lambda^2$\cite{srednicki,callan} but also universal characteristics. Are there any universal features in entanglement entropy of the effective field theory?, perhaps a consequence of underlying symmetries of the high energy sector?. In the resonant case we showed that in the asymptotic long time limit the reduced density matrix becomes the thermal density matrix of a (nearly) free field at the (common) temperature of the heavy fields. We extracted the leading terms dominated by the resonance but neglected higher order terms which would require higher loop contributions to the effective action. Are these higher order terms also ultraviolet divergent?.

\textbf{4:} We focused on obtaining the entanglement entropy in the asymptotic long time limit.  The time evolution of entropy production is   of interest, however it would probably require a rather intense numerical study for a given spectral density and parameters, certainly beyond the scope of this article. Such study would inform on the \emph{rate} of entanglement entropy production\cite{giddings}. In the non-resonant case it is likely that such rate would depend on the initial conditions since as discussed above the effective temperature depends on time as a consequence of squeezing and the memory of the initial conditions even at asymptotically long time. In the resonant case, this rate is very likely $\Gamma_k$ which is the relaxation rate towards equilibrium although this conjecture should be scrutinized further.

\textbf{5:} In the resonant case we found that the light field thermalizes at the same (common) temperature of the heavy degrees of freedom yielding a thermal entropy, which is larger than the entropy of the initial state. Unlike the non-resonant case, in this case the details of the coupling to the heavy sector have been ``erased'' as the thermal state does not reveal any feature of this interaction. This is the result of the sharp resonance in the weak coupling limit and approximating the propagator by a narrow Breit-Wigner form. However, we expect that including width effects there will be corrections to the thermal density matrix  which are perturbatively small in the weak coupling case but which nonetheless may lead to distortions of the thermal spectrum and hint to non-universal details of the interactions in the effective field field theory encoded in the entropy. Such corrections had been recently reported in a condensed matter setting\cite{boyjas}, the study of this possible corrections within the realm of effective field theory would be a worthy endeavor.

We expect to report on some of these issues in forthcoming studies.

\acknowledgements The author   gratefully   acknowledges  support from NSF through grant PHY-1506912. He also thanks D. Jasnow   for enlightening conversations.

\appendix

\section{Spectral density}\label{app:specdens}
The spectral density for the case of two real bosonic fields at a common temperature $T$ was derived in ref.\cite{qbmboy} where the reader is referred to for details. We summarize here the final form, it is given by
\be  \rho_{ld}(q_0,q;T) =  \frac{\mathrm{sign}(q_0)}{8\pi \beta q}\Bigg\{\ln\Big[ \frac{1}{1-e^{-\beta \xi(q_0,q)}}\Big]  + M_1 \leftrightarrow  M_2 \Bigg\}\label{rhold}\ee
\be \rho_d(q_0,q;T) = - \frac{\mathrm{sign}(q_0)}{8\pi \beta q}\Bigg\{\ln\Bigg[ \frac{1-e^{-\beta w_+}}{1-e^{-\beta w_-}}\Bigg]+ M_1 \leftrightarrow  M_2    \Bigg\}\label{rhode}\ee
\bea \rho_{2p}(q_0,q;T) & = &   \frac{\mathrm{sign}(q_0)}{8\pi Q^2}\Bigg\{ \Big[Q^2-(M_1-M_2)^2 \Big]~\Big[Q^2-(M_1+M_2)^2 \Big]  \Bigg\}^{\frac{1}{2}} \nonumber \\ & + & \frac{\mathrm{sign}(q_0)}{8\pi\beta q} \Bigg\{\ln\Bigg[ \frac{1-e^{-\beta w_+}}{1-e^{-\beta w_-}} \Bigg]+ M_1 \leftrightarrow  M_2\Bigg\} \label{rho2chitos} \eea where
\be \xi(q_0,q) = \frac{1}{2|Q^2|}\Bigg\{|q_0| \, \alpha + q \sqrt{\alpha^2+4|Q^2|M^2_1}\Bigg\}\,. \label{xis}\ee
\be \alpha = Q^2 + M^2_1-M^2_2 ~~;~~ \alpha^2-4Q^2M^2_1 = \Big[Q^2-(M_1-M_2)^2 \Big]~\Big[Q^2-(M_1+M_2)^2 \Big]\,. \label{alfas}\ee and
\be w_{\pm}(q_0,q) = \frac{1}{2Q^2} \Bigg\{|q_0|\, \alpha \pm q\,\sqrt{\alpha^2-4Q^2M^2_1} \Bigg\} \label{wpm} \ee

\section{Coefficients}\label{sec:coeffs}

The various integrals needed are of the form
\be I = \int^t_0 dt_1 \int^t_0 dt_2 G_a(t-t_1) G_b(t-t_2) \mathcal{N}_k (t_1-t_2) \label{int}\ee where
$G_{a,b}(t-t_1)$ are proportional to either the Green's function (\ref{resgoft}) or its time derivative, and $\mathcal{N}_k(t_1-t_2)$ is given by (\ref{noiset}). It is convenient to introduce
\be \widetilde{\rho}(p_0,k) = \frac{g^2}{2}\,\rho(p_0,k)\,\mathrm{coth\Big[ \frac{ p_0}{2 T}\Big]} \label{widerho}\,,\ee we note that $\widetilde{\rho}(-p_0,k) = \widetilde{\rho}(p_0,k)$. Now introduce $\tau_{1,2} = t-t_{1,2}$, and take the asymptotic long time limit $t \rightarrow \infty$ in the upper limit of the integrals. One type of integrals required is
\be F_k\equiv  \int^\infty_0 d\tau_1 \int^\infty_0 d\tau_2 \,e^{-\gamma \tau_1}\,e^{-\gamma \tau_2}\, \frac{\sin(\Omega_k \tau_1)}{\Omega_k}\,\frac{\sin(\Omega_k \tau_2)}{\Omega_k}~\mathcal{N}_k(\tau_2-\tau_1)\,,\label{funcF}\ee where $\gamma = \Gamma_k/2$. Using the representation (\ref{noiset}) along with the definition (\ref{widerho}), it is straightforward to find
\be F_k = \int \frac{dp_0}{2\pi}\, \frac{\widetilde{\rho}(p_0,k)}{\Big[\gamma^2 + (p_0+\Omega_k)^2 \Big] \Big[\gamma^2 + (p_0-\Omega_k)^2 \Big]} \,.\label{Ffina}\ee Similarly, we find
\be \int^\infty_0 d\tau_1 \int^\infty_0 d\tau_2 \,e^{-\gamma \tau_1}\,e^{-\gamma \tau_2}\, \cos(\Omega_k \tau_1) \,\frac{\sin(\Omega_k \tau_2)}{\Omega_k}~\mathcal{N}_k(\tau_2-\tau_1) = \gamma \, F_k \,. \label{intcosin}\ee

\be \int^\infty_0 d\tau_1 \int^\infty_0 d\tau_2 \,e^{-\gamma \tau_1}\,e^{-\gamma \tau_2}\, \cos(\Omega_k \tau_1) \,\cos(\Omega_k \tau_2)~\mathcal{N}_k(\tau_2-\tau_1) =  \int \frac{dp_0}{2\pi}\, \frac{\widetilde{\rho}(p_0,k)\,(\gamma^2+p^2_0)}{\Big[\gamma^2 + (p_0+\Omega_k)^2 \Big] \Big[\gamma^2 + (p_0-\Omega_k)^2 \Big]}  \,. \label{intcoscos}\ee The coefficients are combinations of these three types of integrals. We note that with $\gamma = \Gamma_k/2 \propto g^2$, to leading order in $g^2$ we can safely neglect the
$\gamma^2$ term in (\ref{intcoscos}).  In terms of these basic integrals we find
\be K_F =  \int \frac{dp_0}{2\pi}\, \frac{p^2_0\,\widetilde{\rho}(p_0,k)  }{\Big[\gamma^2 + (p_0+\Omega_k)^2 \Big] \Big[\gamma^2 + (p_0-\Omega_k)^2 \Big]}\label{KF}\ee
\be J_1 = \frac{F_k}{\mathcal{G}^2_k(t)} \,,\label{J1} \ee
\be J_2 = \int \frac{dp_0}{2\pi}\, \frac{ \widetilde{\rho}(p_0,k)\,\Big[ \Big(\gamma-\Omega_k \, \frac{\cos(\Omega_k t)}{\sin(\Omega_k t)}\Big)^2 + p^2_0 \Big]}{\Big[\gamma^2 + (p_0+\Omega_k)^2 \Big] \Big[\gamma^2 + (p_0-\Omega_k)^2 \Big]}\,,   \label{J2} \ee
\be J_3 = \frac{\Big(\gamma-\Omega_k \, \frac{\cos(\Omega_k t)}{\sin(\Omega_k t)}\Big)}{\mathcal{G}_k(t)}\,F_k \,,\label{J3}\ee
\be \eta_I = \frac{F_k}{\mathcal{G}_k(t)}\,, \label{etaI}\ee
\be \eta_F = -\Omega_k \, \frac{\cos(\Omega_k t)}{\sin(\Omega_k t)}\,F_k \,.\label{etaF}\ee

%\input{infoqftbiblio}

%\begin{thebibliography}{999}

%\end{thebibliography}


\begin{thebibliography}{999}

\bibitem{ma} S. K. Ma, \emph{Modern theory of critical phenomena}, W. A. Benjamin, Inc. Reading, Mass. 1976.


    \bibitem{fradkin} E. Fradkin, \emph{Field theories of condensed matter physics}, Cambridge Univ. Press. NY. 2013.

   \bibitem{nicolis} S. Dubovsky, L. Hui, A. Nicolis, D. T. Son,  Phys.Rev.\textbf{D85}, 085029 (2012);  A. Nicolis,  arXiv:1108.2513.

  \bibitem{polo} Sašo Grozdanov, Janos Polonyi,   	Phys. Rev. \textbf{D 91}, 105031 (2015).

   \bibitem{eft1} H. Georgi, Annu. Rev. Nucl. Part. Sci.  \textbf{43} 209, (1993).

\bibitem{eft2} A. Pich,  	arXiv:hep-ph/9806303.

\bibitem{eft3} I. Z. Rothstein,  arXiv:hep-ph/0308266.

\bibitem{eft4} C. Burgess, Ann. Rev. Nucl. Part. Sci. \textbf{57}, 329 (2007), Living Rev. Rel. 7 (2004) 5 [gr-qc/0311082];  arXiv:1711.10592.

    \bibitem{dawson} S. Dawson,  	arXiv:1712.07232.

\bibitem{weinberg}  S. Weinberg, Phys.Rev.\textbf{D77}, 123541 (2008).

\bibitem{cheung} C. Cheung, P. Creminelli, A. L. Fitzpatrick, J Kaplan, L. Senatore,  	JHEP 0803, 014 (2008).

\bibitem{senatore} J. J. M. Carrasco, M. P. Hertzberg, L. Senatore,  JHEP, Volume 2012, Number 9 (2012), 82; R. A. Porto, L. Senatore, M. Zaldarriaga,  JCAP 1405 (2014) 022; JHEP
1204 (2012) 024; D. L. Nacir, R. A. Porto, L. Senatore, M. Zaldarriaga,  JHEP \textbf{1201}, 075 (2012).

\bibitem{nuestroreviu} D. Boyanovsky, C. Destri, H. J. de Vega, N. G. Sanchez,  Int.J.Mod.Phys.\textbf{A24},3669 (2009).


    \bibitem{gold} N. Goldenfeld,\emph{ Lectures on Phase Transitions and the Renormalization group},
    (Addison Wesley, Mass. 1992).

     \bibitem{feyver} R.P Feynman and F. L. Vernon, Ann. Phys. (N.Y.)
\textbf{24}, 118 (1963).

\bibitem{leggett} A. O. Caldeira and A. J.
Leggett, Physica \textbf{A 121}, 587 (1983); Phys. Rev. Lett. \textbf{46}, 211 (1981); Annals of Physics, \textbf{149}, 374, (1983).

\bibitem{breuer} N. P. Breuer, F. Petruccione, \emph{The theory of open quantum systems}, Oxford University Press, Oxford, 2007.

\bibitem{zoller} C. Gardiner, P. Zoeller, \emph{Quantum Noise} Springer-Verlag, Berlin (2010).

\bibitem{qinfo} M. A. Nielsen, I. L. Chuang, \emph{Quantum computation and quantum information} (Cambridge University Press, Cambridge, 2000).

\bibitem{calhubuk} E. Calzetta, B.-L. Hu, \emph{Nonequilibrium Quantum Field Theory}, (Cambridge Monographs on Mathematical Physics) Cambridge University Press, Cambridge, 2008.

   \bibitem{calhu} E. Calzetta, B. L. Hu, arXiv:hep-th/9501040; E. Calzetta and B. L. Hu, Phys. Rev. \textbf{D37}, 2878, (1988);
E. Calzetta, B. L. Hu, Phys. Rev. \textbf{D40}, 656 (1989); E. Calzetta, B-L. Hu,  Phys.Rev.\textbf{D55},3536 (1997).

\bibitem{flemhu} C. H. Fleming, B. L. Hu,  Annals of Physics \textbf{327}, 1238 (2012).

 \bibitem{qbmboy} D. Boyanovsky, New J. Phys. \textbf{17}   063017 (2015).

 \bibitem{boycosmo} D. Boyanovsky,  Phys. Rev. \textbf{D 92}, 023527 (2015);  Phys. Rev. \textbf{D 93}, 043501 (2016); Phys. Rev. \textbf{D 93}, 083507 (2016).

  \bibitem{holman} C.P. Burgess, R. Holman, G. Tasinato,  JHEP 1601, 153   (2016).

  \bibitem{shandera} S. Shandera, N. Agarwal, A. Kamal,  arXiv:1708.00493.

  \bibitem{hollowood} T. J. Hollowood, J. I. McDonald, Phys. Rev. \textbf{D95},103521 (2017).

\bibitem{lim} E. A. Lim, Phys. Rev. \textbf{D91}, 083522 (2015).

 \bibitem{braaten} E. Braaten, H.-W. Hammer, G. Peter Lepage,  Phys. Rev. \textbf{D 94}, 056006 (2016);   arXiv:1612.08047.

\bibitem{deboni} D. De Boni,     	 JHEP 1708, 064 (2017).

\bibitem{agon} C. Agon, V. Balasubramanian, S. Kasko, A. Lawrence, arXiv:1412.3148.

\bibitem{bala} V. Balasubramanian, M. B. McDermott, M. Van Raamsdonk, Phys.Rev. \textbf{D86} 045014, (2012).

    \bibitem{teresi} D. Teresi, G. Compagno, arXiv: 1012.3915.

\bibitem{harlow} D. Harlow, Rev. Mod. Phys. \textbf{88} 015002 (2016).

\bibitem{ford} G. W. Ford, R. F.O'Connell,  	Physica\textbf{ E 29}, 82 (2005).

\bibitem{jar} A. Bartolotta, S. Deffner, arXiv:1710.00829.

\bibitem{sorkin} L. Bombelli, R. K. Koul, J. Lee,  R. D. Sorkin
Phys. Rev.\textbf{ D 34}, 373 (1986).

\bibitem{srednicki} M. Srednicki, Phys. Rev. Lett. \textbf{71}, 666 (1993).

\bibitem{callan} C. Callan, F. Wilczek, Phys. Lett. \textbf{B333}, 55, (1994).

\bibitem{cardy} P. Calabrese, J. Cardy,  J.Stat.Mech. 0406 (2004) P06002.

\bibitem{neu} Th. M. Nieuwenhuizen, A. E. Allahverdyan,  Phys. Rev. \textbf{E 66}, 036102 (2002).














   \bibitem{schwinger} J. Schwinger, J. Math. Phys. \textbf{2}, 407
(1961).

 \bibitem{keldysh} L. Keldysh,  Zh.Eksp.Teor.Fiz. \textbf{47}, 1515
(1964)

 \bibitem{maha} P. M. Bakshi and K. T. Mahanthappa,  J.Math.Phys. \textbf{4}1 (1963), J.Math.Phys.\textbf{ 4} 12 (1963).





\bibitem{grabert} H. Grabert, P. Schramm and G.-L. Ingold, Phys.
Rept. \textbf{168}, 115 (1988).

\bibitem{boylan} D. Boyanovsky, H. J. de Vega, R. Holman, S. Prem Kumar, R. D. Pisarski, Phys.Rev.\textbf{D58}, 125009 (1998); D. Boyanovsky, H. J. de Vega, R. Holman, S. Prem Kumar, Rob D. Pisarski, J. Salgado,  arXiv:hep-ph/9810209; D. Boyanovsky, M. D'attanasio, H.J. de Vega, R. Holman, D.-S.Lee,  Phys.Rev. \textbf{D52}, 6805 (1995).

 \bibitem{lellonu} L. Lello, D. Boyanovsky, R. D. Pisarski,    Phys. Rev. \textbf{D 95}, 043524 (2017).

\bibitem{GR} I. S. Gradshteyn, I. M Ryzhik Table of Integrals, Series and Products, 7th Ed.
(Academic Press, Elsevier, Amsterdam, 2007), (see formula 7.37.8).

\bibitem{feynman} R. P. Feynman, \emph{Statistical Mechanics, A set of Lectures}, (Addison Wesley, California, 1972).


  \bibitem{zubairy} M. O. Scully, M. S. Zubairy, \emph{Quantum Optics} (Cambridge University Press,
Cambridge, 1997).

\bibitem{giddings} S. B. Giddings, M. Rota,  arXiv:1710.00005.

\bibitem{lello1} L. Lello, D. Boyanovsky, R. Holman,   JHEP \textbf{1311}, 116 (2013).

\bibitem{seki} R. Peschanski, S. Seki,  Phys.Lett. \textbf{B758}, 89 (2016); I.Y. Park, S. Seki, S.-J. Sin,  Phys.Lett.\textbf{ B743 }, 147 (2015).

\bibitem{semenoff} G. Grignani, G. W. Semenoff,       Phys.Lett. \textbf{B772}, 699  (2017).

\bibitem{lelloentcos} L. Lello, D. Boyanovsky, R. Holman, JHEP\textbf{1404} 055 (2014).

\bibitem{boyjas} D. Boyanovsky, D. Jasnow,  Phys. Rev.\textbf{ A 96}, 062108 (2017).



\end{thebibliography}
\end{document}